\documentclass[
prl
,twocolumn
,nofootinbib
,floatfix
,superscriptaddress
]{revtex4-2} 

\usepackage[utf8]{inputenc} 
\UseRawInputEncoding 
\usepackage{subcaption}
\usepackage[export]{adjustbox}
\usepackage{wrapfig}

\usepackage{notes2bib}
\usepackage{natbib}
\usepackage{gensymb}
\usepackage{amsthm}
\usepackage{amsmath}
\usepackage{color}
\usepackage{bm}
\usepackage{amsmath,amssymb}
\usepackage{amsfonts}
\usepackage{dsfont}
\usepackage{graphicx}
\usepackage{float}
\usepackage{verbatim}
\usepackage{amsfonts}
\usepackage{bbold}
\usepackage{dcolumn}
\usepackage{bm}
\usepackage[dvipsnames]{xcolor}
\usepackage{listings}
\definecolor{blueprl}{RGB}{46,48,146}
\usepackage[pdftex,colorlinks=true,urlcolor=blueprl,citecolor=blueprl,linkcolor=blueprl]{hyperref}
\usepackage{mathrsfs}
\usepackage{mathtools}
\usepackage{times}
\usepackage{color}
\usepackage[dvipsnames]{xcolor}
\usepackage{tikz}
\usepackage{upgreek}

\usepackage{braket}

\usepackage{soul}

\usepackage{mathtools}
\usepackage{dsfont}
\usepackage[mathscr]{euscript}
\usepackage{amsmath}
\usepackage{graphicx}
\usepackage{dcolumn}
\usepackage{bm}
\usepackage{epsfig}
\usepackage{amssymb,latexsym,mathrsfs}
\usepackage{graphicx}
\usepackage{color}
\usepackage{hyperref}
\usepackage{float}
\usepackage{diagbox}
  \usepackage{cancel}
  \usepackage[normalem]{ulem}

\def\beq{\begin{equation}}
\def\eeq{\end{equation}}
\def\bea{\begin{eqnarray}}
\def\eea{\end{eqnarray}}

\definecolor{vividviolet}{rgb}{0.62, 0.0, 1.0}
\definecolor{amaranth}{rgb}{0.9, 0.17, 0.31}
\definecolor{palatinateblue}{rgb}{0.15, 0.23, 0.89}
\definecolor{brightpink}{rgb}{1.0, 0.0, 0.5}
\definecolor{cornflowerblue}{rgb}{0.39, 0.58, 0.93}
\definecolor{deepcarminepink}{rgb}{0.94, 0.19, 0.22}
\definecolor{radicalred}{rgb}{1.0, 0.21, 0.37}
\definecolor{blueblue}{RGB}{21,47,181}
\definecolor{greengreen}{RGB}{65,166,16}
\definecolor{curveMone}{rgb}{0.24,0.60,0.80}
\definecolor{curveMtwo}{rgb}{0.95,0.627,0.1425}
\definecolor{curveMthree}{rgb}{0.455,0.70,0.21}
\definecolor{curveMfour}{rgb}{0.922526,0.385626,0.209179}
\definecolor{curveMfive}{rgb}{0.578,0.51,0.85}
\newcommand{\mnlegend}{{\large%
\textcolor{curveMone}{\rule[0.55ex]{1.4em}{0.22em}}\,$M_1$\quad
\textcolor{curveMtwo}{\rule[0.55ex]{1.4em}{0.22em}}\,$M_2$\quad
\textcolor{curveMthree}{\rule[0.55ex]{1.4em}{0.22em}}\,$M_3$\quad
\textcolor{curveMfour}{\rule[0.55ex]{1.4em}{0.22em}}\,$M_4$\quad
\textcolor{curveMfive}{\rule[0.55ex]{1.4em}{0.22em}}\,$M_5$}}

 \definecolor{darkraspberry}{rgb}{0.53, 0.15, 0.34}

\usepackage{xtab,afterpage}
\usepackage{hyperref}
\usepackage{setspace,calc}

\allowdisplaybreaks

\hypersetup{ linktoc=all,
    colorlinks, linkcolor={blueblue},
    citecolor={greengreen}, urlcolor={blueblue}
}

\newcommand{\be}{\begin{equation}}
\newcommand{\ee}{\end{equation}}
\newcommand{\bs}{\begin{split}} 

\newsavebox{\myhbar}

\begin{document}

\title{Quantum superposition of Cosmic String spacetimes}
\author{Enrique Arias}
\email{earias@iprj.uerj.br}
\affiliation{Instituto Politécnico, Universidade do Estado do Rio de Janeiro, Nova Friburgo - RJ, CEP 28.625-570, Brazil.}
\author{Shoshauna Gauvin}
\email{s3gauvin@uwaterloo.ca}
\affiliation{Department of Physics and Astronomy, University of Waterloo, Waterloo, Ontario, Canada, N2L 3G1}
\author{Willy Izquierdo}
\email{willy.izquierdo@unesp.br}
\affiliation{Instituto de Física Teórica, Universidade Estadual Paulista, São Paulo - SP, CEP 01140-070, Brazil}
\author{Robert B.\ Mann}
\email{rbmann@uwaterloo.ca}
\affiliation{Department of Physics and Astronomy, University of Waterloo, Waterloo, Ontario, Canada, N2L 3G1}
\affiliation{Perimeter Institute, 31 Caroline St., Waterloo, Ontario, N2L 2Y5, Canada}

\begin{abstract}
Since the quantum nature of spacetime is anticipated by a theory of Quantum Gravity, the possibility of spacetime superposition would be a natural consequence.
This spacetime interference phenomena could bring us some clues about what effects could be expected over quantum fields and their interaction with matter. 
In this article, we investigate the effect of the superposition of a Cosmic String spacetime on the Unruh-De Witt detector. This is of interest as the spacetimes possess flat geometry but non-trivial topology. We analyze the effect of the interference of a Cosmic String with different topological charge over the Unruh-De Witt detector response function to vacuum fluctuations. 
We obtain some quantum interference effects, due to the superposition of spacetimes, which would be absent if a classical probabilistic mixture of spacetimes would be considered. In this manner, we obtain real physical consequences that would be indications of a Quantum Gravity phenomena.
\end{abstract} 

 
\maketitle 

\section{Introduction}

Cosmic strings are topological defects that could arise in the early stages of the primordial universe. Many grand unification theory predict cosmic strings as a natural consequence. They form as vortices of the Higgs field that could appear after some symmetry-breaking phase transition \cite{Kibble1,Kibble2,NIELSEN197345,cosmic_string_book}. 
They have significant gravitational lensing properties and could be a primary source of ultrahigh frequency gravitational waves \cite{Servant2024,Kibble2}. 
Cosmic strings also provide a window into quantum gravity via string theory insofar as
they could have been produced in the early universe and stretched to macroscopic scales \cite{Witten:1985fp}, though there is no evidence of this at present. 

The quest for a definitive quantum theory of gravity is still an unfinished task. Although the two dominant proposals, string theory and loop quantum gravity, have led to very interesting results and new theoretical perspectives, a conclusive experimental-based verification is still lacking \cite{grav_decohe_BLHU2022}.  Indeed, obtaining a measurable quantum property of spacetime is a key objective for any theory of quantum gravity.
From this perspective, indications of quantum gravitational phenomena in table-top experiments are an interesting line of research.  This   \textit{bottom-up} approach embraces a phenomenological study of quantum gravity. By analyzing mesoscopic quantum systems and measuring microscopic gravity effects, this line of research aims at obtaining definitive results about the interplay between quantum theory and gravitation \cite{CHRISTODOULOU201964,Alelmeyer2012,Alpelmeyer2018,quant_simu_RevModPhy2014}.
 Within this approach, we could implement an operational method to look for the implications of quantum gravity phenomena, as spacetime quantum superposition. 

A generic phenomenological expectation of any quantum theory of gravity is that of placing spacetime in a state of superposition, with the accompanying interference effects between the different spacetime states. 
Recently a method for 
investigating detectable  signatures of spacetimes in quantum-controlled superpositions of different configurations
was proposed \cite{Foo2021}. This approach  is grounded in quantum field theoretic methods in curved spacetime, and  considers the evolution of a two-level system coupled to the quantum spacetime through its interaction with an ambient quantum field. It has
 subsequently been explored for its possible implications for the quantum nature of spacetime \cite{CHRISTODOULOU201964,RMann2022,Foo2021,RMann2023}. 
 Superposed black hole spacetimes have been of particular interest;  UDW detectors were employed to obtain quantum signatures of black hole mass superposition, with results commensurate with an extension of Bekenstein's conjecture \cite{RMann2022}.
Quantum interference due to the superposition of de Sitter spaces have also been studied \cite{Foo2021}, and even superpositions of  flat Minkowski spacetime with different periodically identified dimensions, have been shown to exhibit non-trivial quantum intereference \cite{RMann2023}.
This approach offers   a unique perspective in understanding the phenomenology of spacetime superposition \cite{dewitt2011role,Anastopoulos_2015,akilsemiclassicalPhysRevD.108.044051}, which has been considered in the context of 
generalizations of quantum reference frames
\cite{delaHamette2023,giacomini10.1116/5.0070018,hoehn2023matter}, decoherence \cite{Arrasmith2019,kieferPhysRevD.53.7050}, and indefinite causal structure \cite{Bell4time, Henderson:2020zax} and
low-energy quantum gravity \cite{Belenchia:2018szb,CHRISTODOULOU201964,Chen2023quantumstatesof,Galley2022nogotheoremnatureof}.

To this end, in this paper we investigate the response of a model detector placed in a superposition of cosmic string spacetimes. 
Cosmic string spacetimes are locally flat and are characterized by a single parameter $\mu$, the linear energy density of the string \cite{PhysRevD.23.852,PhysRevD.31.3288,cosmic_string_book,anderson2015mathematical,Fulling_2012}. They 
have very interesting non-trivial topological properties.  The spacetime around a cosmic string is characterized by a deficit angle \cite{PhysRevD.23.852,PhysRevD.31.3288,anderson2015mathematical,marder1959}.
The effects of this axial topological structure on quantum fields can be described using the  method of images \cite{Linet1987,FROLOV_PhysRevD.35.3779,PhysRevD.34.1918,Fulling_2012}.

In a cosmic string background, quantum fields can develop many interesting properties due to this non-trivial topology. For example, the quantum vacuum state can exhibit Casimir-Polder zero-point energy mechanical effects \cite{Linet1987,FROLOV_PhysRevD.35.3779,BezerradeMello:2011nv,saharian_kotanjyan_2011,PhysRevD.34.1918,Fulling_2012}. 
An Unruh--DeWitt (UDW) detector probes the field two-point function and therefore provides an operational way to distinguish the cosmic-string vacuum from Minkowski vacuum; its transition probability differs from the Minkowski case even for inertial trajectories \cite{Unruh76,DeWitt:1980hx,Svaiter_1994,PCWDavies_1988,bilge_hortacsu_ozdemir_1998,PhysRevD.51.2591}. Recently it has been shown that vacuum entanglement harvesting can be enhanced in a cosmic string background \cite{reznik_2003,Reznik:2003mnx,Pozas-Kerstjens:2015gta,Mart_n_Mart_nez_2016,Ji:2024fcq,Izquierdo2025,huang_he_2020,he_yu_hu_2020,huang_2021}.

 The effects of a superposition of cosmic strings will, amongsst other possiblities, be manifest in a  quantum vacuum.  
Here we seek to understand 
the resultant
observational signatures via UDW detectors and the subsequent implications for quantum gravity.

We employ units $\hbar=c=G=1$. The physical string tension
$\mu_{\rm phys}$ determines $b=1-4\mu_{\rm phys}$ and $\delta=8\pi\mu_{\rm phys}$
\cite{PhysRevD.23.852,PhysRevD.31.3288,anderson2015mathematical}.
We restrict attention to integer quotients, $N=1/b\in\mathbb N$.
Here topological charge denotes this quotient order.

The branch labels $|N\rangle$ and $|M\rangle$ denote geometries
with angular periods $2\pi/N$ and $2\pi/M$, respectively.
Their tensions are $\mu_{{\rm phys},L}=(1-1/L)/4$, for $L=N,M$.
The notation $|\mu_i\rangle$ below labels these geometries,
the angular periods are not their physical tensions.

\section{Quantum field theory on cosmic string spacetime}

The spacetime for an infinite straight cosmic string in the $z$-direction has a metric given by
\begin{equation}\label{metric}
ds^{2} =dt^{2} -dr^{2} -r^{2} d\varphi ^{2} -dz^{2},
\end{equation}
where the spatial domains are $0<r<\infty$, $-\infty<z<\infty$ and $0<\varphi<2\pi b$. This spacetime is characterized by the parameter $b$, which is less than one and defines a deficit angle in space. If $\mu$ is the linear energy density of the cosmic string then \cite{PhysRevD.23.852,PhysRevD.31.3288,cosmic_string_book,anderson2015mathematical}
$$
b = 1 - 4G\mu = 1-4\,\mu_{\rm phys}
$$ 
    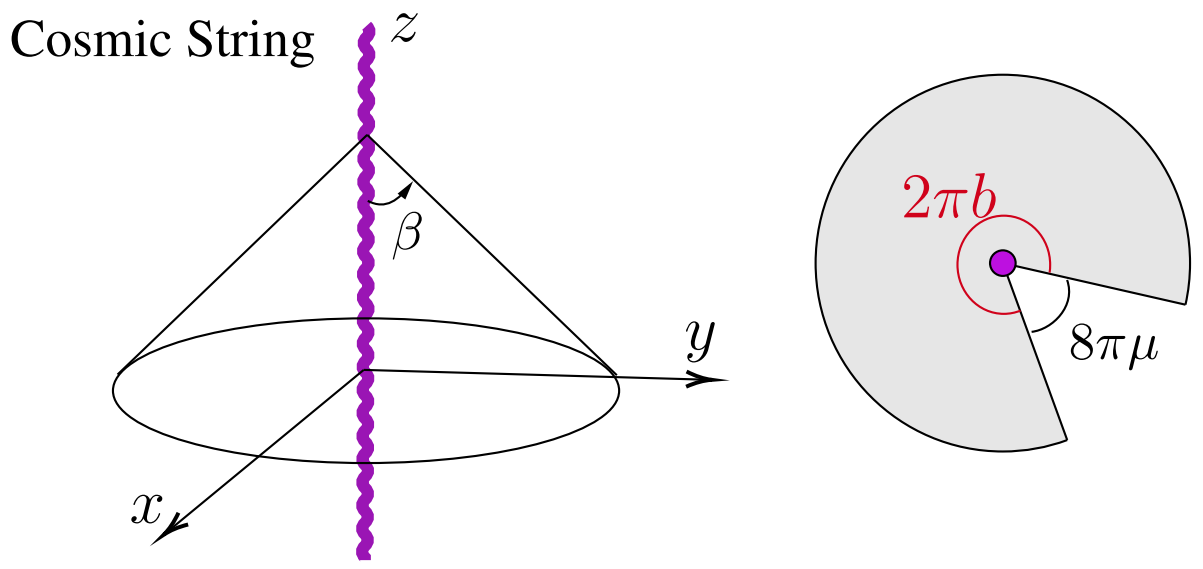
\begin{figure}[t!]
    \centering
\begin{tikzpicture}[x=0.75pt,y=0.75pt,yscale=-.58,xscale=.58]
\draw [color={rgb, 255:red, 154; green, 21; blue, 180 }  ,draw opacity=1 ][fill={rgb, 255:red, 182; green, 36; blue, 211 }  ,fill opacity=1 ][line width=2.25]    (428.74,17.76) .. controls (430.4,19.43) and (430.39,21.1) .. (428.72,22.76) .. controls (427.05,24.43) and (427.05,26.09) .. (428.71,27.76) .. controls (430.37,29.43) and (430.36,31.1) .. (428.69,32.76) .. controls (427.02,34.43) and (427.02,36.09) .. (428.68,37.76) .. controls (430.34,39.43) and (430.33,41.1) .. (428.66,42.76) .. controls (426.99,44.42) and (426.98,46.09) .. (428.64,47.76) .. controls (430.3,49.43) and (430.3,51.09) .. (428.63,52.76) .. controls (426.96,54.42) and (426.95,56.09) .. (428.61,57.76) .. controls (430.27,59.43) and (430.27,61.09) .. (428.6,62.76) .. controls (426.93,64.42) and (426.92,66.09) .. (428.58,67.76) .. controls (430.24,69.43) and (430.23,71.1) .. (428.56,72.76) .. controls (426.89,74.43) and (426.89,76.09) .. (428.55,77.76) .. controls (430.21,79.43) and (430.2,81.1) .. (428.53,82.76) .. controls (426.86,84.43) and (426.86,86.09) .. (428.52,87.76) .. controls (430.18,89.43) and (430.17,91.1) .. (428.5,92.76) .. controls (426.83,94.42) and (426.82,96.09) .. (428.48,97.76) .. controls (430.14,99.43) and (430.14,101.09) .. (428.47,102.76) .. controls (426.8,104.42) and (426.79,106.09) .. (428.45,107.76) .. controls (430.11,109.43) and (430.11,111.09) .. (428.44,112.76) .. controls (426.77,114.42) and (426.76,116.09) .. (428.42,117.76) .. controls (430.08,119.43) and (430.07,121.1) .. (428.4,122.76) .. controls (426.73,124.43) and (426.73,126.09) .. (428.39,127.76) .. controls (430.05,129.43) and (430.04,131.1) .. (428.37,132.76) .. controls (426.7,134.43) and (426.7,136.09) .. (428.36,137.76) .. controls (430.02,139.43) and (430.01,141.1) .. (428.34,142.76) .. controls (426.67,144.42) and (426.66,146.09) .. (428.32,147.76) .. controls (429.98,149.43) and (429.98,151.09) .. (428.31,152.76) .. controls (426.64,154.42) and (426.63,156.09) .. (428.29,157.76) .. controls (429.95,159.43) and (429.95,161.09) .. (428.28,162.76) .. controls (426.61,164.42) and (426.6,166.09) .. (428.26,167.76) .. controls (429.92,169.43) and (429.91,171.1) .. (428.24,172.76) .. controls (426.57,174.43) and (426.57,176.09) .. (428.23,177.76) .. controls (429.89,179.43) and (429.88,181.1) .. (428.21,182.76) .. controls (426.54,184.43) and (426.54,186.09) .. (428.2,187.76) .. controls (429.86,189.43) and (429.85,191.1) .. (428.18,192.76) .. controls (426.51,194.42) and (426.5,196.09) .. (428.16,197.76) .. controls (429.82,199.43) and (429.82,201.09) .. (428.15,202.76) .. controls (426.48,204.42) and (426.47,206.09) .. (428.13,207.76) .. controls (429.79,209.43) and (429.79,211.09) .. (428.12,212.76) .. controls (426.45,214.42) and (426.44,216.09) .. (428.1,217.76) .. controls (429.76,219.43) and (429.75,221.1) .. (428.08,222.76) .. controls (426.41,224.43) and (426.41,226.09) .. (428.07,227.76) .. controls (429.73,229.43) and (429.72,231.1) .. (428.05,232.76) .. controls (426.38,234.43) and (426.38,236.09) .. (428.04,237.76) .. controls (429.7,239.43) and (429.69,241.1) .. (428.02,242.76) .. controls (426.35,244.42) and (426.34,246.09) .. (428,247.76) -- (428,248.93) -- (428,248.93) ;
\draw   (319,175.63) .. controls (319,158.35) and (368.02,144.33) .. (428.5,144.33) .. controls (488.98,144.33) and (538,158.35) .. (538,175.63) .. controls (538,192.92) and (488.98,206.93) .. (428.5,206.93) .. controls (368.02,206.93) and (319,192.92) .. (319,175.63) -- cycle ;
\draw    (321,168.63) -- (429,64.93) ;
\draw    (429,64.93) -- (537,168.93) ;
\draw    (427.5,166.63) -- (342.54,236.66) ;
\draw [shift={(341,237.93)}, rotate = 320.5] [color={rgb, 255:red, 0; green, 0; blue, 0 }  ][line width=0.75]    (10.93,-3.29) .. controls (6.95,-1.4) and (3.31,-0.3) .. (0,0) .. controls (3.31,0.3) and (6.95,1.4) .. (10.93,3.29)   ;
\draw    (427.5,166.63) -- (576,170.88) ;
\draw [shift={(578,170.93)}, rotate = 181.64] [color={rgb, 255:red, 0; green, 0; blue, 0 }  ][line width=0.75]    (10.93,-3.29) .. controls (6.95,-1.4) and (3.31,-0.3) .. (0,0) .. controls (3.31,0.3) and (6.95,1.4) .. (10.93,3.29)   ;
\draw  [draw opacity=0][fill={rgb, 255:red, 0; green, 0; blue, 0 }  ,fill opacity=0.1 ] (731.86,196.98) .. controls (723.17,200.19) and (713.79,201.93) .. (704,201.93) .. controls (659.26,201.93) and (623,165.44) .. (623,120.43) .. controls (623,75.42) and (659.26,38.93) .. (704,38.93) .. controls (748.74,38.93) and (785,75.42) .. (785,120.43) .. controls (785,126.61) and (784.32,132.63) .. (783.02,138.42) -- (704,120.43) -- cycle ; \draw   (731.86,196.98) .. controls (723.17,200.19) and (713.79,201.93) .. (704,201.93) .. controls (659.26,201.93) and (623,165.44) .. (623,120.43) .. controls (623,75.42) and (659.26,38.93) .. (704,38.93) .. controls (748.74,38.93) and (785,75.42) .. (785,120.43) .. controls (785,126.61) and (784.32,132.63) .. (783.02,138.42) ;  
\draw    (704,120.43) -- (731.86,196.98) ;
\draw    (704,120.43) -- (783.02,138.42) ;
\draw    (716.45,150.17) .. controls (725.45,149.17) and (735.8,138.92) .. (731.8,127.72) ;
\draw  [fill={rgb, 255:red, 189; green, 16; blue, 224 }  ,fill opacity=1 ] (698.46,120.43) .. controls (698.46,117.38) and (700.94,114.9) .. (704,114.9) .. controls (707.06,114.9) and (709.54,117.38) .. (709.54,120.43) .. controls (709.54,123.49) and (707.06,125.97) .. (704,125.97) .. controls (700.94,125.97) and (698.46,123.49) .. (698.46,120.43) -- cycle ;
\draw  [draw opacity=0] (711.65,140.97) .. controls (709.41,141.88) and (706.98,142.38) .. (704.43,142.38) .. controls (693.36,142.38) and (684.38,132.88) .. (684.38,121.15) .. controls (684.38,109.42) and (693.36,99.91) .. (704.43,99.91) .. controls (715.51,99.91) and (724.48,109.42) .. (724.48,121.15) .. controls (724.48,122.31) and (724.4,123.44) .. (724.23,124.55) -- (704.43,121.15) -- cycle ; \draw  [color={rgb, 255:red, 208; green, 2; blue, 27 }  ,draw opacity=1 ] (711.65,140.97) .. controls (709.41,141.88) and (706.98,142.38) .. (704.43,142.38) .. controls (693.36,142.38) and (684.38,132.88) .. (684.38,121.15) .. controls (684.38,109.42) and (693.36,99.91) .. (704.43,99.91) .. controls (715.51,99.91) and (724.48,109.42) .. (724.48,121.15) .. controls (724.48,122.31) and (724.4,123.44) .. (724.23,124.55) ;  
\draw    (429.25,93.5) .. controls (437.3,97.67) and (442.7,92.85) .. (447.67,86.51) ;
\draw [shift={(448.87,84.94)}, rotate = 126.98] [fill={rgb, 255:red, 0; green, 0; blue, 0 }  ][line width=0.08]  [draw opacity=0] (8.4,-2.1) -- (0,0) -- (8.4,2.1) -- cycle    ;
\draw (272.33,13) node [anchor=north west][inner sep=0.75pt]   [align=left] {Cosmic String};
\draw (324,218.73) node [anchor=north west][inner sep=0.75pt]  [font=\large]  {$x$};
\draw (564,143.73) node [anchor=north west][inner sep=0.75pt]  [font=\large]  {$y$};
\draw (436,10.4) node [anchor=north west][inner sep=0.75pt]  [font=\large]  {$z$};
\draw (730.29,144.85) node [anchor=north west][inner sep=0.75pt]    {$8\pi \mu $};
\draw (657.8,79.29) node [anchor=north west][inner sep=0.75pt]  [font=\large,color={rgb, 255:red, 208; green, 2; blue, 27 }  ,opacity=1 ]  {$2\pi b$};
\draw (437.2,96.27) node [anchor=north west][inner sep=0.75pt]  [color={rgb, 255:red, 0; green, 0; blue, 0 }  ,opacity=1 ]  {$\beta $};
\end{tikzpicture}
\caption{Cosmic string conical structure and deficit angle.}
    \label{cs_space}
\end{figure}
with $G$ Newton's constant. 
The deficit angle around a cosmic string is given by $\delta=8\pi G\mu $, and this gives a conical structure to this spacetime with a semi-angle $\beta=\arcsin b$ \cite{marder1959,Fulling_2012}, as shown in 
figure~\ref{cs_space}.  The stress energy of the string that yields the metric
\eqref{metric} is
\begin{equation}
    T^0_0 = T^z_z = \mu_{\rm phys}\delta_\perp^{(2)}
    \qquad T^r_r=T^\varphi_\varphi = 0
\end{equation}
The transverse delta distribution is normalized by
$\int_0^\infty r\,dr\int_0^{2\pi b}d\varphi\,
\delta_\perp^{(2)}=1$. Upon integration over the $(r,\varphi)$  yields
the tension $\mu$, or energy/length of the string.  
Henceforth we set $G=1$.

Consider a massless quantum scalar field in this background. It satisfies the Klein-Gordon equation  \cite{Birrell,Parker,Fulling,WaldQFTCST}
\begin{equation}
\label{Klein-Gordon with founts equation for cosmic string}
    \left(\partial^2_t-K\right)\phi(t,{\bf x})=0,
\end{equation} 
where   the spatial operator $K$ in cylindrical coordinates is
\begin{equation}\label{operator for spectral transformation}
    K=\partial^2_r+\frac{1}{r}\partial_r+\frac{1}{r^2}\partial^2_\varphi+\partial^2_z.
\end{equation}
In order to find the quantum propagators or the Green functions of the theory, we introduce the spectral transformation of the operator $K$, whose  characteristic equation is\begin{equation}\label{eigenfuncion equation}
    K\psi_{{\bf k}}({\bf x})=-\omega^2_{{\bf k}}\psi_{{\bf k}}({\bf x}),
\end{equation} where $\omega^2_{\textbf{k}}$ are the eigenvalues and $\psi_{\textbf{k}}$ are orthonormal eigenfunctions.
To solve the eigenvalue problem in    
\eqref{eigenfuncion equation} we expand $\psi_{{\bf k}}(r,\varphi,z)$ in  a Fourier series  since it is periodic in the angular degree of freedom $\varphi$, with period $2\pi b$. Using   separation of variables it is straightforward to show that the normalized solutions are given by
\begin{equation}\label{autofuntion of K operator2}
    \psi_{{\bf k}}(r,\varphi,z)=\sqrt{\frac{p}{2\pi b}}J_{|\eta\, l|}(p\,r)e^{i  \eta\, l\,\varphi}e^{i \kappa z},
\end{equation} 
 where $\eta=1/b$ and  $J_l$ is the usual Bessel function of order $l$. The solutions 
are parametrized by $p\in(0,\infty)$, $\kappa\in(-\infty,\infty)$
and indexed by $l=\{0,\pm1,\pm2,...\}$. 
In the massless field case, the eigenvalues are given by
 $\omega^2_{\textbf{k}}=p^2+\kappa^2$. 
Using these eigenfunctions, we can construct the Wightman positive frequency correlation function  
\begin{equation}\label{Wight1}
G^{+} (x,x')=\int d\Sigma({\bf k})\frac{e^{-i\omega _{k} (t-t')}}{2\omega _{k}} \psi _{{\bf k}} ({\bf x} )\psi _{{\bf k}}^{*} (\mathbf{x} '),
\end{equation}
where 
 \begin{equation}
    \int d\Sigma({\bf k})=\sum^\infty_{l=-\infty}\int_{-\infty}^{\infty} \frac{d\kappa}{2\pi}\int^\infty_0 \!\!\!dp 
\end{equation}
is the appropriate measure of the spectral transformation. 

\begin{figure}[t!]
    \centering
   \tikzset{every picture/.style={line width=0.75pt}} 
\begin{tikzpicture}[x=0.75pt,y=0.75pt,yscale=-.8,xscale=.8]
\draw   (174,165.48) .. controls (174,95.63) and (230.63,39) .. (300.48,39) .. controls (370.33,39) and (426.96,95.63) .. (426.96,165.48) .. controls (426.96,235.33) and (370.33,291.96) .. (300.48,291.96) .. controls (230.63,291.96) and (174,235.33) .. (174,165.48) -- cycle ;
\draw    (122.48,165.48) -- (476.48,165.48) ;
\draw [shift={(478.48,165.48)}, rotate = 180] [color={rgb, 255:red, 0; green, 0; blue, 0 }  ][line width=0.75]    (10.93,-3.29) .. controls (6.95,-1.4) and (3.31,-0.3) .. (0,0) .. controls (3.31,0.3) and (6.95,1.4) .. (10.93,3.29)   ;
\draw    (299.96,329.41) -- (298.97,9.41) ;
\draw [shift={(298.96,7.41)}, rotate = 89.82] [color={rgb, 255:red, 0; green, 0; blue, 0 }  ][line width=0.75]    (10.93,-3.29) .. controls (6.95,-1.4) and (3.31,-0.3) .. (0,0) .. controls (3.31,0.3) and (6.95,1.4) .. (10.93,3.29)   ;
\draw [color={rgb, 255:red, 61; green, 30; blue, 241 }  ,draw opacity=1 ]   (338.99,165.3) .. controls (344.84,160.06) and (342.43,155.96) .. (338.61,149.04) ;
\draw [shift={(337.65,147.3)}, rotate = 61.56] [color={rgb, 255:red, 61; green, 30; blue, 241 }  ,draw opacity=1 ][line width=0.75]    (10.93,-3.29) .. controls (6.95,-1.4) and (3.31,-0.3) .. (0,0) .. controls (3.31,0.3) and (6.95,1.4) .. (10.93,3.29)   ;
\draw  [fill={rgb, 255:red, 245; green, 166; blue, 35 }  ,fill opacity=1 ] (407.58,106.41) .. controls (407.58,103.44) and (409.99,101.03) .. (412.96,101.03) .. controls (415.93,101.03) and (418.34,103.44) .. (418.34,106.41) .. controls (418.34,109.38) and (415.93,111.79) .. (412.96,111.79) .. controls (409.99,111.79) and (407.58,109.38) .. (407.58,106.41) -- cycle ;
\draw  [fill={rgb, 255:red, 245; green, 166; blue, 35 }  ,fill opacity=1 ] (365.27,62.3) .. controls (365.27,59.33) and (367.68,56.92) .. (370.65,56.92) .. controls (373.63,56.92) and (376.04,59.33) .. (376.04,62.3) .. controls (376.04,65.27) and (373.63,67.68) .. (370.65,67.68) .. controls (367.68,67.68) and (365.27,65.27) .. (365.27,62.3) -- cycle ;
\draw [color={rgb, 255:red, 245; green, 166; blue, 35 }  ,draw opacity=1 ]   (336.99,145.55) .. controls (337.59,136.27) and (332.73,134.85) .. (327.11,132.44) ;
\draw [shift={(325.32,131.63)}, rotate = 25.35] [color={rgb, 255:red, 245; green, 166; blue, 35 }  ,draw opacity=1 ][line width=0.75]    (10.93,-3.29) .. controls (6.95,-1.4) and (3.31,-0.3) .. (0,0) .. controls (3.31,0.3) and (6.95,1.4) .. (10.93,3.29)   ;
\draw  [fill={rgb, 255:red, 245; green, 166; blue, 35 }  ,fill opacity=1 ] (305.27,40.63) .. controls (305.27,37.66) and (307.68,35.25) .. (310.65,35.25) .. controls (313.63,35.25) and (316.04,37.66) .. (316.04,40.63) .. controls (316.04,43.61) and (313.63,46.02) .. (310.65,46.02) .. controls (307.68,46.02) and (305.27,43.61) .. (305.27,40.63) -- cycle ;
\draw  [fill={rgb, 255:red, 245; green, 166; blue, 35 }  ,fill opacity=1 ] (236.6,53.3) .. controls (236.6,50.33) and (239.01,47.92) .. (241.99,47.92) .. controls (244.96,47.92) and (247.37,50.33) .. (247.37,53.3) .. controls (247.37,56.27) and (244.96,58.68) .. (241.99,58.68) .. controls (239.01,58.68) and (236.6,56.27) .. (236.6,53.3) -- cycle ;
\draw  [fill={rgb, 255:red, 245; green, 166; blue, 35 }  ,fill opacity=1 ] (417.94,192.25) .. controls (417.94,189.28) and (420.35,186.87) .. (423.32,186.87) .. controls (426.29,186.87) and (428.7,189.28) .. (428.7,192.25) .. controls (428.7,195.22) and (426.29,197.63) .. (423.32,197.63) .. controls (420.35,197.63) and (417.94,195.22) .. (417.94,192.25) -- cycle ;
\draw  [fill={rgb, 255:red, 245; green, 166; blue, 35 }  ,fill opacity=1 ] (385.27,254.25) .. controls (385.27,251.28) and (387.68,248.87) .. (390.65,248.87) .. controls (393.63,248.87) and (396.04,251.28) .. (396.04,254.25) .. controls (396.04,257.22) and (393.63,259.63) .. (390.65,259.63) .. controls (387.68,259.63) and (385.27,257.22) .. (385.27,254.25) -- cycle ;
\draw  [fill={rgb, 255:red, 245; green, 166; blue, 35 }  ,fill opacity=1 ] (330.6,286.58) .. controls (330.6,283.61) and (333.01,281.2) .. (335.99,281.2) .. controls (338.96,281.2) and (341.37,283.61) .. (341.37,286.58) .. controls (341.37,289.56) and (338.96,291.97) .. (335.99,291.97) .. controls (333.01,291.97) and (330.6,289.56) .. (330.6,286.58) -- cycle ;
\draw [color={rgb, 255:red, 29; green, 19; blue, 241 }  ,draw opacity=1 ][line width=0.75]  [dash pattern={on 4.5pt off 4.5pt}]  (300.48,165.48) -- (310.49,42.63) ;
\draw [shift={(310.65,40.63)}, rotate = 94.66] [color={rgb, 255:red, 29; green, 19; blue, 241 }  ,draw opacity=1 ][line width=0.75]    (9.84,-2.96) .. controls (6.25,-1.25) and (2.97,-0.27) .. (0,0) .. controls (2.97,0.27) and (6.25,1.26) .. (9.84,2.96)   ;
\draw [color={rgb, 255:red, 29; green, 19; blue, 241 }  ,draw opacity=1 ][line width=0.75]  [dash pattern={on 4.5pt off 4.5pt}]  (300.48,165.48) -- (241.27,55.39) ;
\draw [shift={(240.32,53.63)}, rotate = 61.73] [color={rgb, 255:red, 29; green, 19; blue, 241 }  ,draw opacity=1 ][line width=0.75]    (9.84,-2.96) .. controls (6.25,-1.25) and (2.97,-0.27) .. (0,0) .. controls (2.97,0.27) and (6.25,1.26) .. (9.84,2.96)   ;
\draw [color={rgb, 255:red, 29; green, 19; blue, 241 }  ,draw opacity=1 ][line width=0.75]  [dash pattern={on 4.5pt off 4.5pt}]  (300.48,165.48) -- (421.37,191.82) ;
\draw [shift={(423.32,192.25)}, rotate = 192.29] [color={rgb, 255:red, 29; green, 19; blue, 241 }  ,draw opacity=1 ][line width=0.75]    (9.84,-2.96) .. controls (6.25,-1.25) and (2.97,-0.27) .. (0,0) .. controls (2.97,0.27) and (6.25,1.26) .. (9.84,2.96)   ;
\draw [color={rgb, 255:red, 29; green, 19; blue, 241 }  ,draw opacity=1 ][line width=0.75]  [dash pattern={on 4.5pt off 4.5pt}]  (301.28,166.08) -- (390.03,253.45) ;
\draw [shift={(391.45,254.85)}, rotate = 224.55] [color={rgb, 255:red, 29; green, 19; blue, 241 }  ,draw opacity=1 ][line width=0.75]    (9.84,-2.96) .. controls (6.25,-1.25) and (2.97,-0.27) .. (0,0) .. controls (2.97,0.27) and (6.25,1.26) .. (9.84,2.96)   ;
\draw [color={rgb, 255:red, 29; green, 19; blue, 241 }  ,draw opacity=1 ][line width=0.75]  [dash pattern={on 4.5pt off 4.5pt}]  (300.48,165.48) -- (334.93,286.51) ;
\draw [shift={(335.48,288.43)}, rotate = 254.11] [color={rgb, 255:red, 29; green, 19; blue, 241 }  ,draw opacity=1 ][line width=0.75]    (9.84,-2.96) .. controls (6.25,-1.25) and (2.97,-0.27) .. (0,0) .. controls (2.97,0.27) and (6.25,1.26) .. (9.84,2.96)   ;
\draw [color={rgb, 255:red, 34; green, 11; blue, 237 }  ,draw opacity=1 ][line width=1.5]  [dash pattern={on 1.69pt off 2.76pt}]  (247.99,127.3) .. controls (219.99,163.3) and (232.99,215.97) .. (280.99,230.97) ;
\draw [color={rgb, 255:red, 245; green, 166; blue, 35 }  ,draw opacity=1 ]   (321.32,132.3) .. controls (319.45,127.63) and (315.25,127.61) .. (306.26,129.27) ;
\draw [shift={(304.32,129.63)}, rotate = 349.05] [color={rgb, 255:red, 245; green, 166; blue, 35 }  ,draw opacity=1 ][line width=0.75]    (10.93,-3.29) .. controls (6.95,-1.4) and (3.31,-0.3) .. (0,0) .. controls (3.31,0.3) and (6.95,1.4) .. (10.93,3.29)   ;
\draw [color={rgb, 255:red, 245; green, 166; blue, 35 }  ,draw opacity=1 ]   (301.99,130.21) .. controls (297.06,124.78) and (290.11,129.12) .. (285.97,131.66) ;
\draw [shift={(284.32,132.63)}, rotate = 331.39] [color={rgb, 255:red, 245; green, 166; blue, 35 }  ,draw opacity=1 ][line width=0.75]    (10.93,-3.29) .. controls (6.95,-1.4) and (3.31,-0.3) .. (0,0) .. controls (3.31,0.3) and (6.95,1.4) .. (10.93,3.29)   ;
\draw [color={rgb, 255:red, 29; green, 19; blue, 241 }  ,draw opacity=1 ][line width=0.75]  [dash pattern={on 4.5pt off 4.5pt}]  (300.48,165.48) -- (369.53,63.95) ;
\draw [shift={(370.65,62.3)}, rotate = 124.22] [color={rgb, 255:red, 29; green, 19; blue, 241 }  ,draw opacity=1 ][line width=0.75]    (9.84,-2.96) .. controls (6.25,-1.25) and (2.97,-0.27) .. (0,0) .. controls (2.97,0.27) and (6.25,1.26) .. (9.84,2.96)   ;
\draw [color={rgb, 255:red, 29; green, 19; blue, 241 }  ,draw opacity=1 ]   (300.48,165.48) -- (411.19,107.34) ;
\draw [shift={(412.96,106.41)}, rotate = 152.29] [color={rgb, 255:red, 29; green, 19; blue, 241 }  ,draw opacity=1 ][line width=0.75]    (10.93,-3.29) .. controls (6.95,-1.4) and (3.31,-0.3) .. (0,0) .. controls (3.31,0.3) and (6.95,1.4) .. (10.93,3.29)   ;
\draw [color={rgb, 255:red, 245; green, 166; blue, 35 }  ,draw opacity=1 ]   (311.22,202.88) .. controls (318.59,203.83) and (320.97,203.7) .. (325.76,194.23) ;
\draw [shift={(326.62,192.48)}, rotate = 115.74] [color={rgb, 255:red, 245; green, 166; blue, 35 }  ,draw opacity=1 ][line width=0.75]    (10.93,-3.29) .. controls (6.95,-1.4) and (3.31,-0.3) .. (0,0) .. controls (3.31,0.3) and (6.95,1.4) .. (10.93,3.29)   ;
\draw [color={rgb, 255:red, 245; green, 166; blue, 35 }  ,draw opacity=1 ]   (328.62,192.08) .. controls (336.29,191.9) and (338.19,185.42) .. (338.4,176.4) ;
\draw [shift={(338.42,174.48)}, rotate = 90] [color={rgb, 255:red, 245; green, 166; blue, 35 }  ,draw opacity=1 ][line width=0.75]    (10.93,-3.29) .. controls (6.95,-1.4) and (3.31,-0.3) .. (0,0) .. controls (3.31,0.3) and (6.95,1.4) .. (10.93,3.29)   ;
\draw (347.93,143.93) node [anchor=north west][inner sep=0.75pt]  [color={rgb, 255:red, 13; green, 27; blue, 202 }  ,opacity=1 ]  {$\varphi $};
\draw (423.07,96.09) node [anchor=north west][inner sep=0.75pt]  [font=\small]  {$n=0$};
\draw (376.47,40.54) node [anchor=north west][inner sep=0.75pt]  [font=\small]  {$n=1$};
\draw (301.27,15.74) node [anchor=north west][inner sep=0.75pt]  [font=\small]  {$n=2$};
\draw (220.87,25.4) node [anchor=north west][inner sep=0.75pt]  [font=\small]  {$n=3$};
\draw (436.47,186.4) node [anchor=north west][inner sep=0.75pt]  [font=\small]  {$n=N-1$};
\draw (398.07,254.8) node [anchor=north west][inner sep=0.75pt]  [font=\small]  {$n=N-2$};
\draw (310.07,297.6) node [anchor=north west][inner sep=0.75pt]  [font=\small]  {$n=N-3$};
\draw (335.27,117.6) node [anchor=north west][inner sep=0.75pt]  [font=\footnotesize,color={rgb, 255:red, 0; green, 0; blue, 0 }  ,opacity=0.88 ]  {$\Delta \upnu $};
\draw (306.93,103.27) node [anchor=north west][inner sep=0.75pt]  [font=\footnotesize]  {$\Delta \upnu $};
\draw (274.93,102.27) node [anchor=north west][inner sep=0.75pt]  [font=\footnotesize]  {$\Delta \upnu $};
\draw (340.93,185.68) node [anchor=north west][inner sep=0.75pt]  [font=\footnotesize]  {$\Delta \upnu $};
\draw (317.27,204.68) node [anchor=north west][inner sep=0.75pt]  [font=\footnotesize]  {$\Delta \upnu $};
\draw (485.2,155.4) node [anchor=north west][inner sep=0.75pt]    {$x$};
\draw (280.4,0.4) node [anchor=north west][inner sep=0.75pt]    {$y$};
\draw (381.2,120.8) node [anchor=north west][inner sep=0.75pt]  [color={rgb, 255:red, 23; green, 47; blue, 240 }  ,opacity=1 ]  {$r$};
\end{tikzpicture}
    \caption{Image structure of the conical spacetime of cosmic string with the topological charge $N$ and $\Delta\upnu=2\pi/N$.}
    \label{image_structure}
\end{figure}
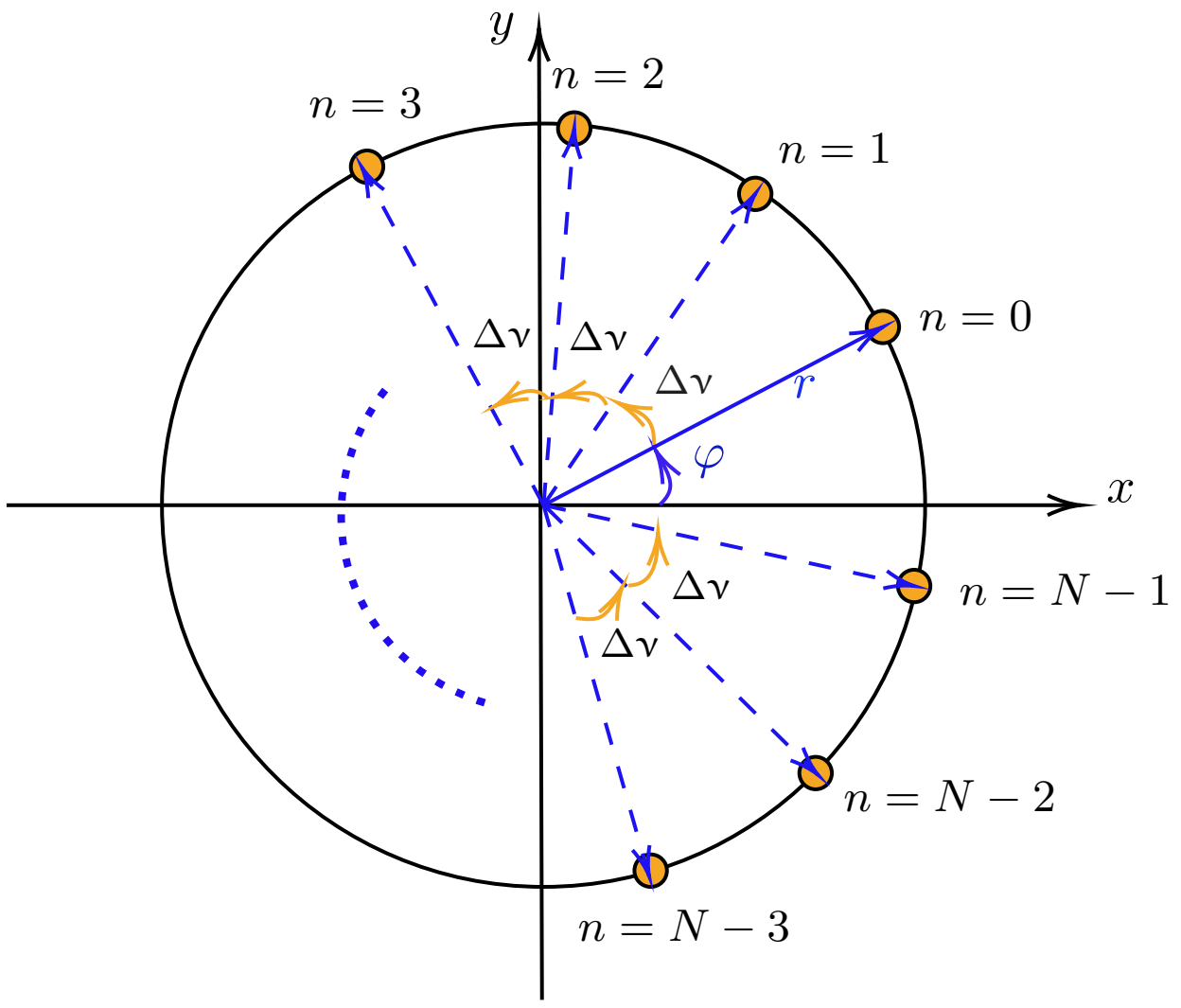

For the special case $\eta=N\in\mathbb{N}$ the 
correlation function \eqref{Wight1}
admits a finite $N$-image description, enabling closed-form expressions via the method of images \cite{Linet1987,FROLOV_PhysRevD.35.3779,PhysRevD.34.1918,Fulling_2012}. (Related classical radiation phenomena in cosmic-string backgrounds also single out integer $\eta$  \cite{ALIEV1989142}.) 
 The integer $N$ is identified as the topological charge of the cosmic string \cite{ALIEV1989142}, and there exists a topological image structure around the cosmic string with $N$ images, as shown in figure~\ref{image_structure}. 

For such topologically charged strings,  
the Wightman function \eqref{Wight1} can be written as an image sum \cite{Linet1987,FROLOV_PhysRevD.35.3779,PhysRevD.34.1918,Fulling_2012}
\begin{widetext}
  \begin{align}
G^{+} (x,x')= & -\frac{1}{4\pi ^{2}}\sum _{k=0}^{N -1}\frac{1}{(\Delta t-i\epsilon )^{2} -\left( \Delta r^{2} +\Delta z^{2} +4rr'\sin^{2} (\pi k/N +\Delta \varphi /2)\right)} ,
\label{Wight2}
\end{align} 
\end{widetext}
where the spacetime points are $x=(t,{\bf x})=(t,r,\varphi,z)$ and $x'=(t',{\bf x}')=(t',r',\varphi',z')$, and $\Delta t=t-t'$, $\Delta r=r-r'$, $\Delta \varphi=\varphi-\varphi'$ and $\Delta z=z-z'$.
The image structure shown in figure~\ref{image_structure} is such that we have $N$ identical wedge-shaped sectors of angular distance $\Delta\upnu=2\pi/N$,
such that for any point $x=(t,r,\varphi,z)$
there are $N-1$ equivalent image points 
$x_n=(t,{\bf x}_n)=(t,r,\varphi_n,z)=(t,r,\varphi+n\Delta\upnu,z)$ 
where   $n=1,...,N-1$. In this way, we see that the Wightman function \eqref{Wight2} can be interpreted as the image sum of the corresponding correlation function \cite{Birrell,Fulling}
\begin{equation}\label{Wight3}
    G^+_0(x,x')=-\frac{1}{4\pi^2}\frac{1}{(\Delta t-i\epsilon)^2-({\bf x}-{\bf x}')^2}
\end{equation}
in Minkowski spacetime between some point $x$  (and all its images) with some other point $x'$ 
\begin{equation*}
G^{+} (x,x')=\sum _{n=0}^{N-1} G_{0}^{+} (x_n ,x')
\end{equation*}
on the cosmic string spacetime.

Noting that the Wightman function in Minkowski spacetime 
only depends on the spatial distance between the points we can rewrite
\eqref{Wight3} as 
\begin{equation}\label{Wight4}
G^{+} (x,x')=\sum _{n=0}^{N-1} G_{0}^{+} ({\cal J}_{N}^{n} x ,x').
\end{equation}
by introducing the discrete rotation isometry  (deck transformation) operator ${\cal J}_N$, defined around a cosmic string with topological charge $N$. 
This operator provides us with the image of a given point in space by performing a counterclockwise rotation by an angle $\Delta\upnu$
\begin{equation*}
   {\cal J}_N:(t,r,\varphi,z)\longrightarrow (t,r,\varphi+\Delta\upnu,z),
\end{equation*}
and so  the $n$-th image of a given point is 
$$
x_n={\cal J}_N^n\; x
$$
namely 
the result of the application of ${\cal J}_N$ $n$-times. 
We can show that the rotation of $n$-times of the first point and $n'$-times of the second corresponds to a total rotation given just by the difference $n-n'$, in such a manner that
$$G_{0}^{+} ({\cal J}_{N}^{n} x ,{\cal J}_{N}^{n'} x')=G_{0}^{+} ({\cal J}_{N}^{n-n'} x ,x').$$
Consequently the Wightman function can be expressed as a double sum over the images of point $x$ and $x'$, with some proper normalization, as follows
\begin{equation}
G^+(x,x')=\sum _{n=0}^{N-1}\sum _{n'=0}^{N-1}\left(\frac{1}{c_{n,n'}^{N}}\right) G^+_0\left( {\cal J}_{N}^{n} x,{\cal J}_{N}^{n'} x'\right),
\end{equation}
where the normalization coefficient is given by
\begin{gather*}
c_{n,n'}^{N} =( 2-\delta _{n,n'})( N-|n-n'|).
\end{gather*}
This expression will be important when we consider  the correlation function in the case of superposed cosmic strings.

\section{Unruh-De Witt detector on quantum superposition of spacetimes}

We now wish to consider the effects of quantum superposition of spacetimes on the response function of a single Unruh  De Witt  (UdW) detector. 

As usual, we model a UdW detector as a two level system with Hamiltonian given by \cite{Unruh76,DeWitt:1980hx,Mart_n_Mart_nez_2016}
\begin{equation}  
H_{D} =\Omega _{g} |g\rangle \langle g|+\Omega _{e} |e\rangle \langle e|,
\end{equation}
besides this detector we have a massless scalar field, $\phi_\mu$, with the usual Klein-Gordon Hamiltonian \cite{Birrell,Parker}
\begin{equation*}
H_{F} =\int d^{3} x\left(\frac{1}{2}( \partial _{t} \phi_\mu )^{2} +\frac{1}{2}( \nabla \phi_\mu )^{2} 
\right),
\end{equation*}
The branch fields defined below are automorphic operators
constructed from $\phi_0$ and inherit their time evolution
from this common free Hamiltonian  \cite{Linet1987,PhysRevD.34.1918,FROLOV_PhysRevD.35.3779}. In an energy basis, the state of the spacetime will be $|\mu \rangle$.

We wish to consider the superposition of two cosmic string states $\displaystyle \{|\mu _{1} \rangle ,|\mu _{2} \rangle \}$. 
If there is interaction between the detector (D), quantum field (F) and spacetime (S) then all the dynamics would be given by a block diagonal free Hamiltonian of the form \cite{Louko_2006,Louko_2008,Satz_2007,Juárez-Aubry_2014}
\begin{equation}
\mathbb{H}_{0} =H_{D} +H_{F} +H_{S},
\end{equation}
where  $H_S$ is the Hamiltonian of the string.
This latter quantity is somewhat subtle. Although the string has a well defined tension $\mu$, its total energy (obtained by integration of $T^t_{\;\nu}\xi^\nu$ over all space, where $\xi^\nu$ is a timelike Killing vector) is formally infinite, due to the infinite length of the string. We therefore interpret $H_S$ as an effective Hamiltonian for a
fiducial segment of length $\ell$:
\[
H_S=\omega_N|N\rangle\langle N|+\omega_M|M\rangle\langle M|,
\qquad
\omega_L=\ell\mu_{\rm phys,L}.
\]
The segment length $\ell$ is fixed independently of the
detector switching width $\mathcal T$. Only the difference $\Delta\omega=\omega_M-\omega_N$ enters the detector response. On a given string state $|\mu \rangle $, the Hamiltonian divided by this length scale yields the tension of the string that gives rise to the spacetime \eqref{metric}. 
 

We denote $\displaystyle H_{D} =H_{D} \otimes I_{F} \otimes I_{S}$, where $\displaystyle I$ are the identity operators in each corresponding Fock space, and analogously for the other terms. We use the Minkowski vacuum of a common covering field. The branch fields are the automorphic operators defined below. 

Employing the same reasoning for constructing interaction Hamiltonians in superposed spacetimes as for the Minkowski \cite{RMann2023},
de Sitter \cite{Foo2021}, and black hole \cite{RMann2022} cases,
a single UdW detector linearly coupled to  the quantum field in a superposition of cosmic strings,  the interaction Hamiltonian is given by
\begin{equation}\label{Hint}
\mathbb{H}_{int} =\lambda \ m_{0}\sum _{i=1}^{2} \phi _{\mu _{i}}( x) |\mu _{i} \rangle \langle \mu _{i} |,
\end{equation}
where  $\lambda$ is the interaction coupling constant, $\phi _{\mu_i }( x)$ is the scalar field representation in the string state $\displaystyle \mu_i$, which is given by the  topological charge.

The UdW detector has spacetime position $\displaystyle x$, and its monopole operator is a combination of the ladder operators,  
\begin{equation}\label{mono1}
m_{0} =   |e\rangle \langle g|+
 |g \rangle \langle e |.
\end{equation}


We write $|0\rangle$ for the Minkowski vacuum of the common
covering field. Let $\rho_{0D}$ denote the initial detector
state and $|\mu\rangle$ the initial string superposition, 
alternatively one can work in $\mathcal{F}_N\oplus\mathcal{F}_M$ and define $G^+_{NM}$ via an isometry between the two Fock spaces. For calculational simplicity we adopt the single-vacuum ansatz here and note it explicitly \cite{Birrell,HOLLANDS20151}.
where $\displaystyle \rho_{0D}$ is the initial density operator of the UdW detector alone, $\displaystyle |0\rangle $ is the (common) vacuum state of the quantum field in both spacetimes, and $\displaystyle |\mu \rangle $ is the initial state of the cosmic string, which we assume is the superposition  
\begin{equation}\label{musup}
|\mu \rangle =c_{1} |\mu _{1} \rangle +c_{2} |\mu _{2} \rangle.
\end{equation}
where $\displaystyle c_{1}$ and $\displaystyle c_{2}$ are complex coefficients that determine the mixture of spacetimes. 
In sum, we assume that the total system (Detector+Field+Spacetime) is in the    state  
\begin{equation}
\rho _{0} =  \rho_{0D}\otimes |0\rangle \langle 0|\otimes |\mu \rangle \langle \mu |,
\end{equation}
at the initial time 
 $\displaystyle t=t_{0}$, 
 where $\rho_{0D}$ is the initial state of the detector.
To obtain the density matrix of the total system at any time, $\displaystyle \rho ( t)$, we need to solve the quantum Liouville equation. 
\begin{equation}
i\frac{\partial \rho ( t)}{\partial t} =[\mathbb{H} ,\rho ( t)],
\end{equation}
where the total Hamiltonian must consider the free and interaction parts, i.e., $\displaystyle \mathbb{H} =\mathbb{H}_{0} +\mathbb{H}_{int}$. In the interaction picture, this becomes
\begin{equation}
i\frac{\partial \rho _{I}( t)}{\partial t} =\left[\mathbb{H}_{int}^{I}( t) ,\rho _{I}( t)\right],
\end{equation}
where we define the density operator in the interaction picture by applying the free time evolution operator 
\begin{equation*}
\mathbf{U}_{0}( t,t_{0}) =\exp( -i\mathbb{H}_{0}( t-t_{0})),
\end{equation*}
where $\displaystyle t_{0}$ is the initial time when the interaction begins, and we have in the interaction picture
\begin{equation}
\rho _{I}( t) =\mathbf{U}_{0}^{\dagger }( t,t_{0}) \ \rho ( t) \ \mathbf{U}_{0}( t,t_{0}).
\end{equation}
The interaction Hamiltonian $\displaystyle \mathbb{H}_{int}$ in the interaction picture is 
\begin{equation*}
\mathbb{H}_{int}^{I} =\mathbf{U}_{0}^{\dagger }( t,t_{0}) \ \mathbb{H}_{int} \ \mathbf{U}_{0}( t,t_{0}) 
\end{equation*}
and since  $\displaystyle \mathbb{H}_{0} =H_{D} +H_{F} +H_{S}$ we can express 
\begin{equation*}
\ \mathbf{U}_{0}( t,t_{0}) =U_{D}( t,t_{0}) \ U_{F}( t,t_{0}) \ U_{S}( t,t_{0})
\end{equation*}
where
\begin{align*}
U_{D}( t,t_{0}) & =\exp( -iH_{D}( t-t_{0})),\\
U_{F}( t,t_{0}) & =\exp( -iH_{F}( t-t_{0})),\\
U_{S}( t,t_{0}) & =\exp( -iH_{S}( t-t_{0})).
\end{align*}
The interaction Hamiltonian in the interaction picture is therefore
\begin{equation}
\mathbb{H}_{int}^{I}( t) =\lambda \chi(t) \ m_{I}( t) \ \sum _{i=1}^{2} \, \phi _{\mu _{i}}^{I}( x) \,|\mu _{i} \rangle \langle \mu _{i} |,
\end{equation}
where $\phi^I(x)$ is the scalar field operator representation in the interaction pictures. It is worth remembering that in this picture, the operators evolve with the free part of the Hamiltonian. The monopole operator \eqref{mono1} in the interaction picture becomes
\begin{align}\label{mono2}
m_{I}( t)
 & =|e\rangle \langle g|\,e^{i\Delta \Omega ( t-t_{0})}+|g\rangle \langle e|\,e^{-i\Delta \Omega ( t-t_{0})},
\end{align}
where $\Delta\Omega=\Omega_e-\Omega_g>0$ is the energy gap of the UDW detector. 

We can solve the Liouville equation in the interaction picture by using perturbation theory.
We can expand in a Dyson series the solution of this equation by assuming a small interaction coupling constant 
\begin{align}
\rho _{I}( t) &=\rho _{0} -i\int _{t_{0}}^{t} d\tau \ \left[\mathbb{H}_{int}^{I}( \tau ) ,\rho _{0}\right] \\\nonumber
&+( -i)^{2}\int _{t_{0}}^{t} d\tau \ \int _{t_{0}}^{\tau } d\tau '\ \left[\mathbb{H}_{int}^{I}( \tau ) ,\left[\mathbb{H}_{int}^{I}( \tau ') ,\rho_{0}\right]\right] +...
\end{align}
Having found the expression for the density operator of the whole system (Detector+Field+Space) by taking the trace over all the possible final states of the quantum field we obtain 
\begin{equation}\label{redmat1}
\widetilde\rho_D(t) =\text{Tr}_{F}( \langle \mu _{f} |\rho _{I}( t) |\mu _{f} \rangle ),
\end{equation}
for the reduced density matrix of the detectors in the superposed cosmic string spacetime. This is the unnormalized postselected detector operator. Its trace is the postselection success probability. The conditional state is:
$\rho_{D|\mu_f}=\widetilde\rho_D/
\operatorname{Tr}_D\widetilde\rho_D$. 

For comparison, the unconditional detector state (no post-selection) is obtained by 
tracing over both field and string states:
\begin{equation}\label{uncond1}
\rho_{\mathrm{uncond}}(t)=\mathrm{Tr}_{F,S}\{\rho_I(t)\}.
\end{equation}
A general final Schr\"odinger-picture measurement state is
$|\xi\rangle=d_1|\mu_1\rangle+d_2|\mu_2\rangle$, with
$|d_1|^2+|d_2|^2=1$. Its interaction-picture representative is
$U_S^\dagger(t,t_0)|\xi\rangle$.
\\
Here we specialize to return postselection, $d_i=c_i$, so that
the measured Schr\"odinger-picture state equals the initially
prepared superposition. This is a choice of final measurement;
the detector, field and string may still become correlated
during the interaction. We obtain
\begin{align}\label{muf}
|\mu _{f} \rangle
 & =c_{1}\,e^{ i\omega _{1}( t-t_{0})}\,|\mu _{1} \rangle +c_{2}\,e^{i\omega _{2}( t-t_{0})}\, |\mu _{2} \rangle.
\end{align}

By considering terms up to second order in perturbation theory, we obtain the final state of the single UdW detector, after the interaction with the quantum vacuum in the presence of superposed cosmic string spacetimes \cite{Mart_n_Mart_nez_2016,Louko_2008,Juárez-Aubry_2014}
\begin{equation}
\widetilde\rho_D(t) = \rho^{( 0)} -i\int _{t_{0}}^{t} \!\!d\tau\,  \rho ^{( 1)} ( \tau ) +( -i)^{2}\int _{t_{0}}^{t}\!\! d\tau \ \int _{t_{0}}^{\tau }\!\! d\tau '\,  \rho ^{( 2)} ( \tau ,\tau ')
\end{equation}
where we define
\begin{align*}
  \rho ^{( 0)}  & =\text{Tr}_{F}( \langle \mu _{f} |\rho _{0} | \mu _{f} \rangle ),\\
  \rho ^{( 1)} ( \tau ) & =\text{ Tr}_{F}\left( \langle \mu _{f} | \ \left[\mathbb{H}_{int}^{I}( \tau ) ,\rho _{0}\right] |\mu _{f} \rangle \right),\\
  \rho ^{( 2)} ( \tau ,\tau ') & =\text{Tr}_{F}\left( \langle \mu _{f} | \ \left[\mathbb{H}_{int}^{I}( \tau ) ,\left[\mathbb{H}_{int}^{I}( \tau ') ,\rho _{0}\right]\right] | \mu _{f} \rangle \right).
\end{align*}
 We can show that the zeroth-order term is given by
\begin{align}\label{rho00}
  \rho ^{( 0)} &= |\langle\mu_f|\mu\rangle|^2 \rho_{0D} \\ 
 & = 
  \rho_{0D} \left( 1-4|c_1|^2|c_2|^2 \ \sin^{2}( \Delta \omega ( t-t_{0}) /2)\right),
  \nonumber
\end{align} 
which is proportional to the initial state $\rho_{0D}$ 
and  the post-selection overlap   $|\langle\mu_f|\mu\rangle|^2$,
where   $\Delta\omega=\omega_2-\omega_1$.

For an initially ground-state detector, dividing the
order-$\lambda^2$ joint excitation probability by
$q_0=|\langle\mu_f|\mu\rangle|^2$ gives its leading conditional
probability. This requires $q_0>0$ and an interaction correction
to the success probability that is small compared with $q_0$.
The exact normalization is
$\operatorname{Tr}_D\widetilde\rho_D=q_0+O(\lambda^2)$;
the leading conditional expression is not uniform near
vanishing $q_0$. 
The first order correction is trivial, i.e. $  \rho ^{( 1)} ( \tau )=0$. The second order correction is somewhat more cumbersome. 
Taking the detector initially in its ground state at $t=t_0$,
\[
\rho_{0D}=|g\rangle\langle g|
=\begin{pmatrix}1&0\\0&0\end{pmatrix}
\]
In the ordered basis $\{|g\rangle,|e\rangle\}$, the normalized
conditional detector state, to the displayed perturbative order, is
\begin{equation}
\rho_{D|\mu_f}(t) =\left(\begin{matrix}
\mathcal{P}_{g} & 0\\
0 & \mathcal{P}_{e}
\end{matrix}\right)
\end{equation}
where the probability of finding the UdW detector at its excited state is given by
\begin{equation}
\mathcal{P}_{e} =\lambda ^{2}\frac{\mathscr{P} ( \Delta \Omega )}{1-4|c_1|^2|c_2|^2\ \sin^{2}( \Delta \omega ( t-t_{0}) /2)},
\end{equation}
and the probability of remaining in the ground state is 
\begin{equation}
\mathcal{P}_{g} =1-\lambda ^{2} \ \frac{\mathscr{P} (\Delta\Omega)}{1-4|c_1|^2|c_2|^2\ \sin^{2}( \Delta \omega ( t-t_{0}) /2)}.
\end{equation}
where  we introduce the function
\begin{align}
\mathscr{P} (E)& =|c_{1} |^{4} F_{11}(E) +|c_{2} |^{4} F_{22}(E)\\
&\!\!\!\!\!\!\!\!\!\!\!+|c_{1} |^{2} |c_{2} |^{2}\left( F_{21}( E) \ e^{i\Delta \omega ( t-t_{0})} +F_{12}(E) \ e^{-i\Delta \omega ( t-t_{0})}\right)
\nonumber
\end{align}
which depends on the response functions  
\begin{align}
F_{ij}(E) = \int_{t_0}^{t}d\tau\int_{t_0}^{t}d\tau'\,
\chi(\tau)\chi(\tau')e^{-iE\Delta\tau}G^+_{ij}(x,x')
\end{align}
where $i,j=\{1,2\}$ with
\begin{equation}\label{Gij}
G^+_{ij}(x,x')=\langle 0|\phi_{\mu_i}(x)\phi_{\mu_j}(x')|0\rangle.  
\end{equation}
For an initial energy eigenstate and a stationary inertial trajectory, the reduced density matrix remains diagonal in the detector energy basis at leading nontrivial order.

We note that correlation functions \eqref{Gij} lead to interference phenomena between the different cosmic string spacetimes. Ultimately, a theory of quantum gravity would give an account of the proper definitions of the correlation functions between different spacetimes. 
The cross-branch correlators $G^+_{12}$ encode genuine quantum coherence between spacetime branches and are therefore theory-dependent. We adopt a simple ansatz motivated by symmetry and consistency requirements --namely, reduction to the standard single-geometry Wightman functions when $N=M$, symmetry under $(N,x)\leftrightarrow(M,x')$, and recovery of the Minkowski correlator when $N=M=1$ -- and use it to quantify the qualitative imprint of spacetime coherence on detector response.

Inspecting  $\mathscr{P} $, we see that the interference terms contribute to the UdW final state only if both coefficients $c_1$ and $c_2$ of the quantum superpositions of spacetimes are non-zero. If one of the $c$-coefficients vanishes, then the interference terms shown in $\mathscr{P}$ vanish and we obtain the response for a single cosmic string spacetime.

Alternatively,  by tracing out both field and string degrees of freedom we obtain the reduced density matrix
\cite{MartinezTjoa2021}
\begin{equation}
\rho_{\mathrm{cl}} ( t) =\text{Tr}_{F}\bigg(\text{Tr}_{S}\big(\rho _{I}(t) \big)\bigg ),
\end{equation}
or
\begin{equation*}
\rho _{\mathrm{cl}}(t)=\left(\begin{matrix}
{\cal P}_{\mathrm{cl}}^g & 0\\
0 & {\cal P}_{\mathrm{cl}}^e
\end{matrix}\right),
\end{equation*}
with 
\begin{align}
\mathscr{P}_{\mathrm{cl}}(E)
= \mathscr{P}_{\mathrm{uncond}}(E)
& =|c_{1} |^{2} F_{11}(E) +|c_{2} |^{2} F_{22}( E),
\end{align}
from   the unconditional (mixed) state \eqref{uncond1},
where   ${\cal P}_{\mathrm{cl}}^e=\lambda^2\mathscr{P}_{\mathrm{cl}}(\Delta\Omega)$.  
 This density matrix is automatically normalized since   ${\cal P}_{\mathrm{cl}}^g=1-{\cal P}_{\mathrm{cl}}^e$.

For clarity, the joint excitation response coefficient for the postselection in Eq.~\eqref{muf} is
\begin{equation}
\begin{split}
\mathscr{P}(E)
&= |c_1|^{4} F_{11}(E) + |c_2|^{4} F_{22}(E)\\
&\quad + 2|c_1|^{2}|c_2|^{2}\,\mathrm{Re}\!\left[e^{-i\Delta\omega\,(t-t_0)}\,F_{12}(E)\right]
\end{split}
\end{equation}
whereas the unconditional excitation response coefficient obtained by tracing over the cosmic string states is  \cite{MartinezTjoa2021}
\begin{equation}
\mathscr{P}_{\mathrm{\mathrm{cl}}}(E)=|c_{1}|^{2}F_{11}(E)+|c_{2}|^{2}F_{22}(E)
\label{eq:uncond-prob}
\end{equation}
where the cross term cancels because $H_{\mathrm{int}}$ is diagonal in the spacetime basis.

\section{Bloch sphere for superposition of spacetimes}

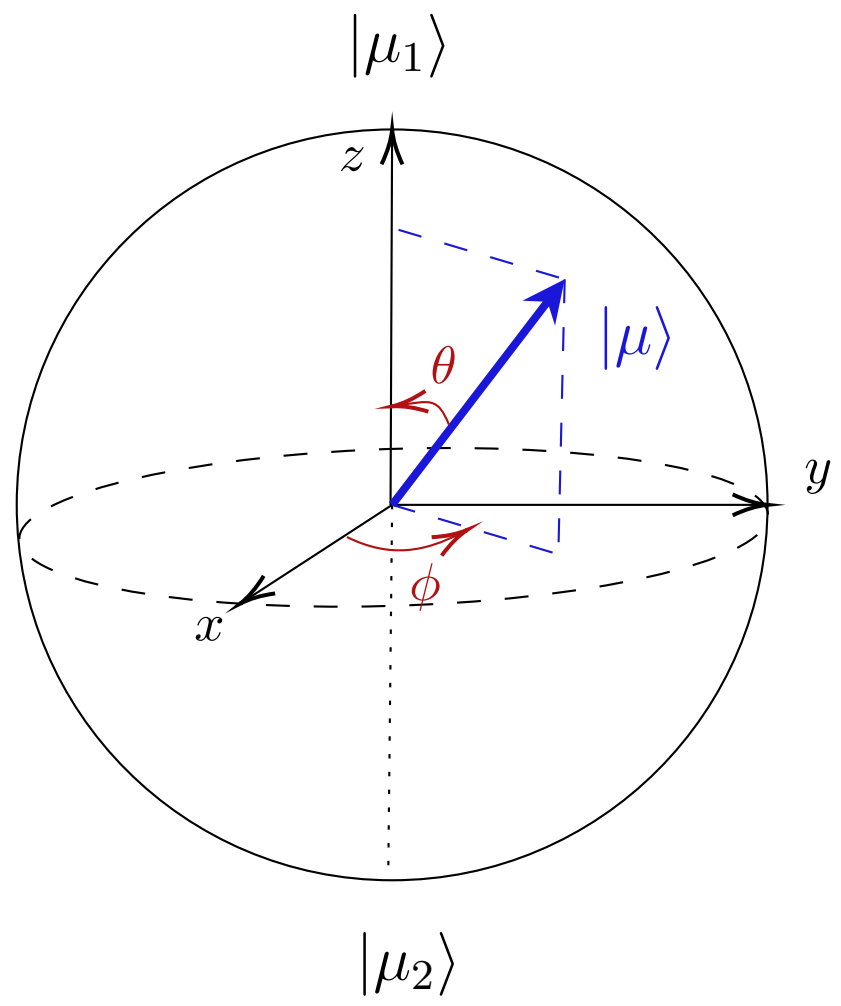
\begin{figure}[t!]
    \centering
    \begin{tikzpicture}[x=0.75pt,y=0.75pt,yscale=-.8,xscale=.8]

\draw    (395,169) -- (510.75,169) ;
\draw [shift={(512.75,169)}, rotate = 180] [color={rgb, 255:red, 0; green, 0; blue, 0 }  ][line width=0.75]    (10.93,-3.29) .. controls (6.95,-1.4) and (3.31,-0.3) .. (0,0) .. controls (3.31,0.3) and (6.95,1.4) .. (10.93,3.29)   ;
\draw    (395,169) -- (349.02,198.9) ;
\draw [shift={(347.35,199.99)}, rotate = 326.96] [color={rgb, 255:red, 0; green, 0; blue, 0 }  ][line width=0.75]    (10.93,-3.29) .. controls (6.95,-1.4) and (3.31,-0.3) .. (0,0) .. controls (3.31,0.3) and (6.95,1.4) .. (10.93,3.29)   ;
\draw    (394.5,169.05) -- (394.99,53.25) ;
\draw [shift={(395,51.25)}, rotate = 90.24] [color={rgb, 255:red, 0; green, 0; blue, 0 }  ][line width=0.75]    (10.93,-3.29) .. controls (6.95,-1.4) and (3.31,-0.3) .. (0,0) .. controls (3.31,0.3) and (6.95,1.4) .. (10.93,3.29)   ;
\draw   (277.25,169) .. controls (277.25,103.97) and (329.97,51.25) .. (395,51.25) .. controls (460.03,51.25) and (512.75,103.97) .. (512.75,169) .. controls (512.75,234.03) and (460.03,286.75) .. (395,286.75) .. controls (329.97,286.75) and (277.25,234.03) .. (277.25,169) -- cycle ;
\draw  [dash pattern={on 4.5pt off 4.5pt}] (347.35,199.99) .. controls (288.2,196.25) and (261.81,182.5) .. (288.41,169.27) .. controls (315,156.05) and (384.5,148.36) .. (443.65,152.11) .. controls (502.79,155.85) and (529.18,169.6) .. (502.59,182.82) .. controls (476,196.05) and (406.49,203.74) .. (347.35,199.99) -- cycle ;
\draw  [dash pattern={on 0.84pt off 2.51pt}]  (395,169) -- (393.8,282.9) ;
\draw [color={rgb, 255:red, 27; green, 23; blue, 217 }  ,draw opacity=1 ][line width=1.5]    (395,169) -- (446.7,101.41) ;
\draw [shift={(449.13,98.24)}, rotate = 127.42] [fill={rgb, 255:red, 27; green, 23; blue, 217 }  ,fill opacity=1 ][line width=0.08]  [draw opacity=0] (13.4,-6.43) -- (0,0) -- (13.4,6.44) -- (8.9,0) -- cycle    ;
\draw [color={rgb, 255:red, 27; green, 23; blue, 217 }  ,draw opacity=1 ] [dash pattern={on 4.5pt off 4.5pt}]  (449.13,98.24) -- (447.13,184.5) ;
\draw [color={rgb, 255:red, 27; green, 23; blue, 217 }  ,draw opacity=1 ] [dash pattern={on 4.5pt off 4.5pt}]  (395,169) -- (447.13,184.5) ;
\draw [color={rgb, 255:red, 27; green, 23; blue, 217 }  ,draw opacity=1 ] [dash pattern={on 4.5pt off 4.5pt}]  (397,82.73) -- (449.13,98.24) ;
\draw [color={rgb, 255:red, 178; green, 16; blue, 18 }  ,draw opacity=1 ]   (380.83,179.09) .. controls (394.43,185.77) and (405.08,183.79) .. (416.9,177.72) ;
\draw [shift={(418.58,176.84)}, rotate = 151.63] [color={rgb, 255:red, 178; green, 16; blue, 18 }  ,draw opacity=1 ][line width=0.75]    (10.93,-3.29) .. controls (6.95,-1.4) and (3.31,-0.3) .. (0,0) .. controls (3.31,0.3) and (6.95,1.4) .. (10.93,3.29)   ;
\draw [color={rgb, 255:red, 178; green, 16; blue, 18 }  ,draw opacity=1 ]   (412.83,143.84) .. controls (409.28,135.33) and (407.3,136.2) .. (396.97,137.8) ;
\draw [shift={(395.08,138.09)}, rotate = 351.7] [color={rgb, 255:red, 178; green, 16; blue, 18 }  ,draw opacity=1 ][line width=0.75]    (10.93,-3.29) .. controls (6.95,-1.4) and (3.31,-0.3) .. (0,0) .. controls (3.31,0.3) and (6.95,1.4) .. (10.93,3.29)   ;

\draw (379,13.4) node [anchor=north west][inner sep=0.75pt]  [font=\large]  {$| \mu _{1} \rangle $};
\draw (382,301.4) node [anchor=north west][inner sep=0.75pt]  [font=\large]  {$| \mu _{2} \rangle $};
\draw (331.33,202.73) node [anchor=north west][inner sep=0.75pt]    {$x$};
\draw (522.67,153.4) node [anchor=north west][inner sep=0.75pt]    {$y$};
\draw (376.67,55.4) node [anchor=north west][inner sep=0.75pt]    {$z$};
\draw (457.67,105.07) node [anchor=north west][inner sep=0.75pt]  [font=\large,color={rgb, 255:red, 27; green, 23; blue, 217 }  ,opacity=1 ]  {$| \mu \rangle $};
\draw (405.42,117.69) node [anchor=north west][inner sep=0.75pt]  [color={rgb, 255:red, 178; green, 16; blue, 18 }  ,opacity=1 ]  {$\theta $};
\draw (398.75,185.7) node [anchor=north west][inner sep=0.75pt]  [color={rgb, 255:red, 178; green, 16; blue, 18 }  ,opacity=1 ]  {$\phi$};
\end{tikzpicture}
    \caption{Bloch sphere for a two-level quantum spacetime.
    }
    \label{fig:enter-label}
\end{figure}

The initial state \eqref{musup} of the spacetime is in a general quantum state located on the Bloch sphere of two possible well-defined spacetime states \cite{Nielsen2000,Steane1998}, as shown in
figure~\ref{fig:enter-label}. 
We can write 
$c_1=\cos(\theta/2)$ and $c_2=e^{i\phi}\sin(\theta/2)$ so that 
\begin{equation}
|\mu\rangle=\cos(\theta/2)\,|\mu_1\rangle+e^{i\phi}\sin(\theta/2)\,|\mu_2\rangle \; .
\end{equation}
This general expression helps us to consider a family of possible spacetime superposition states \cite{Nielsen2000,Steane1998}, since $0<\theta<\pi$ and $0<\phi<2\pi$.
This Bloch-sphere parametrization is purely kinematical: it follows from $\mathrm{span}\{|\mu_1\rangle,|\mu_2\rangle\}\cong\mathbb{C}^2$. If the branch Hamiltonian has nondegenerate eigenvalues $H_S|\mu_i\rangle=\omega_i|\mu_i\rangle$, then the relative phase is dynamical and evolves as $\phi(t)=\phi(t_0)+(\omega_2-\omega_1)(t-t_0)$ under free evolution. Our post-selection state in Eq.~\eqref{muf} precisely accounts for this phase accumulation in the interaction picture.

The visibility depends on the relative dynamical phase accumulated between the two geometric branches during the interaction window, controlled by the dimensionless product $\Delta\omega\,(t-t_0)$. We shall find it convenient to define the dimensionless quantity
\begin{equation}\label{muNM}
\Xi_{NM}:=2\pi\!\left(\frac{1}{N}-\frac{1}{M}\right)=8\pi\Delta\mu,
\end{equation}
where $N$ and $M$ are  the respective topological charges for
the states $|\mu_1\rangle$ and $|\mu_2\rangle$. 
Note that $\Delta\omega= \alpha\,\Xi_{NM}$, where
$\alpha=\ell/(8\pi)$, since
$\Delta\omega=\ell\Delta\mu_{\rm phys}$
and $\Xi_{NM}=8\pi\Delta\mu_{\rm phys}$.

\begin{figure}[b!]
    \centering
\begin{tikzpicture}[x=0.75pt,y=0.75pt,yscale=-.6,xscale=.6]
\draw  [color={rgb, 255:red, 0; green, 0; blue, 0 }  ,draw opacity=1 ][line width=0.75]  (354.33,218.91) .. controls (354.33,191.34) and (429.42,169) .. (522.04,169) .. controls (614.66,169) and (689.74,191.34) .. (689.74,218.91) .. controls (689.74,246.47) and (614.66,268.81) .. (522.04,268.81) .. controls (429.42,268.81) and (354.33,246.47) .. (354.33,218.91) -- cycle ;
\draw [color={rgb, 255:red, 0; green, 0; blue, 0 }  ,draw opacity=1 ][line width=0.75]    (521.07,10.81) -- (355.07,211.48) ;
\draw [color={rgb, 255:red, 0; green, 0; blue, 0 }  ,draw opacity=1 ][line width=0.75]    (521.07,10.81) -- (689.07,212.81) ;
\draw  [color={rgb, 255:red, 27; green, 23; blue, 217 }  ,draw opacity=1 ][line width=0.75]  (430.41,214.81) .. controls (430.41,199.9) and (472.04,187.81) .. (523.41,187.81) .. controls (574.77,187.81) and (616.41,199.9) .. (616.41,214.81) .. controls (616.41,229.72) and (574.77,241.81) .. (523.41,241.81) .. controls (472.04,241.81) and (430.41,229.72) .. (430.41,214.81) -- cycle ;
\draw [color={rgb, 255:red, 27; green, 23; blue, 217 }  ,draw opacity=1 ][line width=0.75]    (521.78,12.67) -- (430.41,214.81) ;
\draw [color={rgb, 255:red, 27; green, 23; blue, 217 }  ,draw opacity=1 ][line width=0.75]    (521.78,12.67) -- (616.41,214.81) ;
\draw [color={rgb, 255:red, 189; green, 16; blue, 224 }  ,draw opacity=1 ][line width=2.25]    (521.78,12.67) .. controls (523.45,14.32) and (523.46,15.99) .. (521.81,17.67) .. controls (520.16,19.34) and (520.17,21.01) .. (521.84,22.67) .. controls (523.52,24.32) and (523.53,25.99) .. (521.88,27.67) .. controls (520.23,29.34) and (520.24,31.01) .. (521.91,32.67) .. controls (523.58,34.33) and (523.59,36) .. (521.94,37.67) .. controls (520.29,39.35) and (520.3,41.02) .. (521.98,42.67) .. controls (523.65,44.33) and (523.66,46) .. (522.01,47.67) .. controls (520.36,49.34) and (520.37,51.01) .. (522.04,52.67) .. controls (523.71,54.33) and (523.72,56) .. (522.07,57.67) .. controls (520.42,59.35) and (520.43,61.02) .. (522.11,62.67) .. controls (523.78,64.33) and (523.79,66) .. (522.14,67.67) .. controls (520.49,69.34) and (520.5,71.01) .. (522.17,72.67) .. controls (523.85,74.32) and (523.86,75.99) .. (522.21,77.67) .. controls (520.56,79.34) and (520.57,81.01) .. (522.24,82.67) .. controls (523.91,84.33) and (523.92,86) .. (522.27,87.67) .. controls (520.62,89.35) and (520.63,91.01) .. (522.31,92.66) .. controls (523.98,94.32) and (523.99,95.99) .. (522.34,97.66) .. controls (520.69,99.33) and (520.7,101) .. (522.37,102.66) .. controls (524.04,104.32) and (524.05,105.99) .. (522.4,107.66) .. controls (520.75,109.34) and (520.76,111.01) .. (522.44,112.66) .. controls (524.11,114.32) and (524.12,115.99) .. (522.47,117.66) .. controls (520.82,119.33) and (520.83,121) .. (522.5,122.66) .. controls (524.18,124.31) and (524.19,125.98) .. (522.54,127.66) .. controls (520.89,129.33) and (520.9,131) .. (522.57,132.66) .. controls (524.24,134.32) and (524.25,135.99) .. (522.6,137.66) .. controls (520.95,139.34) and (520.96,141.01) .. (522.64,142.66) .. controls (524.31,144.32) and (524.32,145.99) .. (522.67,147.66) .. controls (521.02,149.33) and (521.03,151) .. (522.7,152.66) .. controls (524.38,154.31) and (524.39,155.98) .. (522.74,157.66) .. controls (521.09,159.33) and (521.1,161) .. (522.77,162.66) .. controls (524.44,164.32) and (524.45,165.99) .. (522.8,167.66) .. controls (521.15,169.33) and (521.16,171) .. (522.83,172.66) .. controls (524.51,174.31) and (524.52,175.98) .. (522.87,177.66) .. controls (521.22,179.33) and (521.23,181) .. (522.9,182.66) .. controls (524.57,184.32) and (524.58,185.99) .. (522.93,187.66) .. controls (521.28,189.34) and (521.29,191.01) .. (522.97,192.66) .. controls (524.64,194.32) and (524.65,195.99) .. (523,197.66) .. controls (521.35,199.33) and (521.36,201) .. (523.03,202.66) .. controls (524.71,204.31) and (524.72,205.98) .. (523.07,207.66) -- (523.07,208.88) -- (523.07,208.88) ;
\draw (476,132.4) node [anchor=north west][inner sep=0.75pt]  [font=\large]  {$N$};
\draw (403,75.73) node [anchor=north west][inner sep=0.75pt]  [font=\large]  {$M$};
\draw (343.67,355.07) node [anchor=north west][inner sep=0.75pt]    {$\textcolor[rgb]{0.11,0.09,0.85}{\Delta \upnu =}\textcolor[rgb]{0.11,0.09,0.85}{\frac{2\pi }{N}}$};
\draw (343,289.07) node [anchor=north west][inner sep=0.75pt]  [color={rgb, 255:red, 0; green, 0; blue, 0 }  ,opacity=1 ]  {$\Delta \upmu =\frac{2\pi }{M}$};
\draw (443.67,357.4) node [anchor=north west][inner sep=0.75pt]  [color={rgb, 255:red, 27; green, 23; blue, 217 }  ,opacity=1 ]  {$J_{N} :( t,r,\varphi ,z) \longrightarrow ( t,r,\varphi {+\Delta \upnu } ,z) .$};
\draw (440.33,288.4) node [anchor=north west][inner sep=0.75pt]  [color={rgb, 255:red, 0; green, 0; blue, 0 }  ,opacity=1 ]  {$J_{M} :( t,r,\varphi ,z) \longrightarrow ( t,r,\varphi +\Delta \upmu , z) .$};
\end{tikzpicture}
    \caption{Superposition of conical structures of cosmic string spacetimes with topological charges $N$ and $M$.}
\label{cs_superposition}
\end{figure}

We could also take the  $\phi$-dependent phase difference factor to depend on the global properties of the spacetimes; however for this study we shall set this phase to zero. 
If the two branches are identical ($N=M$), then the cross correlator reduces to the ordinary correlator, $G^+_{NM}=G^+_{N}$, and the conditional and unconditional detector responses reduce to the standard single-string result (up to the overall post-selection factor, which is removed when we normalize conditional probabilities) \cite{RMann2023,RMann2022}.

If $N=M$ that $\mathcal{L}_{NN}(E) = \mathcal{F}_N(E)$ and the cross correlator reduces to the ordinary correlator, $G^+_{NM}=G^+_{N}$. In this case, the interference phase becomes trivial, and the conditional and unconditional detector responses reduce to the single-string result (up to the overall post-selection normalization factor).

\section{Green's function for superposition of Cosmic String Spacetimes}

We next consider the correlation functions of the quantum field between different spacetimes.

We begin with the image method that helps us build the Wightman function on a single spacetime \cite{Linet1987,FROLOV_PhysRevD.35.3779,PhysRevD.34.1918,Fulling_2012}. We depict the situation for two superposed cosmic string spacetimes in figure~\ref{cs_superposition}.  The cosmic string state $|\mu_1\rangle$ has topological charge $N$ while the second state $|\mu_2\rangle$ has topological charge $M$.
These spacetimes have a holomorphic operator related to their image topological structure. Let us denote these operators by $J_N$ and $J_M$ respectively \cite{Fulling_2012}. The correlation functions for each spacetime are  
\begin{align}
    G^+_N(x,x')=G^+_{11}(x,x')&=\langle 0|\phi_{\mu_1}(x)\phi_{\mu_1}(x')|0\rangle,\nonumber\\
     G^+_M(x,x')=G^+_{22}(x,x')&=\langle 0|\phi_{\mu_2}(x)\phi_{\mu_2}(x')|0\rangle.
\end{align}
By employing the method of images we obtain  \cite{Linet1987,PhysRevD.34.1918,BezerradeMello:2011nv,Fulling_2012}
\begin{equation}
G_{L}^+(x,x')=\sum _{l=0}^{L-1}\sum _{l'=0}^{L-1}\left(\frac{1}{c_{l,l'}^{L}}\right) G^+_0\left( {\cal J}_{L}^{l} x,{\cal J}_{L}^{l'} x'\right),
\end{equation}
where
\begin{align}
c_{l,l'}^{L} &=L\,( 2-\delta _{l,l'})\left( 1-\left|\frac{l}{L}-\frac{l'}{L}\right|\right),
\end{align}
and we have $L=\{N,M\}$.
The equivalence of the two weighting prescriptions follows
by grouping cyclically identical image terms. Define
$g_k=G_0^+(x,{\cal J}_L^k x')$. Rotational invariance and
${\cal J}_L^L=1$ imply
\[
G_0^+({\cal J}_L^l x,{\cal J}_L^{l'}x')
=g_{(l'-l)\bmod L}.
\]
For each nonzero cyclic class $k=1,\ldots,L-1$, there
are $L-k$ pairs with $l'-l=k$ and $k$ pairs with
$l'-l=k-L$. Since
$c^L_{l,l'}=(2-\delta_{l,l'})(L-|l-l'|)$,
their combined coefficient is
\[
\frac{L-k}{2(L-k)}+\frac{k}{2k}=1.
\]
The diagonal class contains $L$ terms of weight $1/L$,
also giving total coefficient one. Under the uniform
prescription, every cyclic class contains $L$ terms
of weight $1/L$. Therefore,
\begin{align*}
G_L^+(x,x')&=\sum_{k=0}^{L-1}g_k\\
&=\frac{1}{L}\sum_{l,l'=0}^{L-1}
G_0^+({\cal J}_L^l x,{\cal J}_L^{l'}x').
\end{align*}
Thus the complete sums agree after grouping identical
image terms, although the individual pair coefficients
are different. We specify the cross correlator using a common Minkowski
covering field $\phi_0$, with aligned image rotations:
\[
\phi_L(x)=\frac{1}{\sqrt L}\sum_{l=0}^{L-1}
\phi_0({\cal J}_L^l x),\qquad L=N,M.
\]
This common-field embedding is an assumption of the model.
Taking the vacuum expectation value gives
\begin{equation}
G^+_{NM}(x,x')=\frac{1}{\sqrt{NM}}
\sum_{n=0}^{N-1}\sum_{m=0}^{M-1}
G^+_0({\cal J}_N^n x,{\cal J}_M^m x')
\end{equation}
The factor $1/\sqrt{NM}$ is the product of the field
normalizations $1/\sqrt N$ and $1/\sqrt M$.
For $N=M=L$, it becomes $1/L$ and reproduces the
single-geometry correlator through the equivalent image
sum above. Taking geometric means of single-geometry
multiplicity coefficients does not, by itself, determine
a cross correlator for $N\ne M$. Here the cross correlator
is fixed by the stated common-covering-field assumption.
\\
\\
Let $d=\gcd(N,M)$ and $Q=\operatorname{lcm}(N,M)$.
The relative rotations have $Q$ distinct values, each
occurring $d$ times. Therefore,
\[
G^+_{NM}=\frac{d}{\sqrt{NM}}G^+_Q,\qquad
{\cal L}_{NM}=\frac{d}{\sqrt{NM}}{\cal F}_Q.
\]
In particular, $G^+_{NN}=G^+_N$.

These correlation functions define the response function in each individual spacetime
\begin{align}
{\cal F}_{L}(E)
&=\int_{\mathbb R}d\tau\int_{\mathbb R}d\tau'\,
\chi(\tau)\chi(\tau')e^{-iE\Delta\tau}G_L^+(x,x')
\end{align}
with $L=\{N,M\}$, while for the response function that takes into account the superposition of spcaetime we use the notation
\begin{align}
{\cal L}_{NM}(E)
&=\int_{\mathbb R}d\tau\int_{\mathbb R}d\tau'\,
\chi(\tau)\chi(\tau')e^{-iE\Delta\tau}G_{NM}^+(x,x')
\end{align}

For the analytic response functions below, we integrate over
the full Lorentzian switching profile. In order to obtain this we introduce a switching function that turns on and off the interaction only for a finite amount of time. We shall use a Lorentzian switching function
\[
\chi(t)=\frac{(\mathcal T/2)^2}
{(t-t_c)^2+(\mathcal T/2)^2}.
\]
This dimensionless Lorentzian has unit peak and full width
at half maximum $\mathcal T$. Its support is the real line,
and $\int_{\mathbb R}\chi(t)\,dt=\pi\mathcal T/2$.
This choice yields the $e^{-|E|\mathcal{T}}$ factors in Eqs.~(\ref{definition of LNM}) and the 
corresponding expression for ${\cal F}_{L}$ after standard contour integration \cite{Sebastianchlicht_2004,Louko_2006,Louko_2008,Satz_2007,Juárez-Aubry_2014}.

We choose a Lorentzian switching for analytic convenience: it allows the time integrals to be evaluated in closed form by contour methods, yielding explicit dependence on $E\mathcal{T}$ (cf.\ Refs.~\cite{Sebastianchlicht_2004,Louko_2006,Louko_2008,Satz_2007,Juárez-Aubry_2014}). While Gaussian switchings are also common, the qualitative features we focus on—suppression at large gap, enhancement for short interaction times, and the presence/absence of cross-branch interference are expected to be robust under smooth, compactly supported or rapidly decaying envelopes. A direct comparison with Gaussian switching is left for future work.
\\
\\
For the full-profile calculation, we prescribe a finite
postselection phase through the interaction-picture state
$\sum_i c_i e^{i\omega_i\mathcal D}|\mu_i\rangle$.
The reference interval $\mathcal D$ is independent of the
switching width. The plotted scans choose
$\mathcal D=\mathcal T$ and fixed $\ell=1$.
\\
\\
The time integrals leading to these response functions follow standard Unruh--DeWitt detector calculations with Lorentzian switching and can be evaluated by contour integration (see, e.g., Refs.~\cite{Sebastianchlicht_2004,Louko_2006,Louko_2008,Satz_2007,Juárez-Aubry_2014}). We therefore quote the resulting closed-form expressions here.
\begin{widetext}
\begin{equation}
{\cal F}_{L}(E) =\frac{\mathcal{T}^{2}}{16} e^{-|E |\mathcal{T}}\sum _{l=0}^{L-1}\frac{1}{4r^{2}\sin^{2}( \pi l/L) +\mathcal{T}^{2}} -\Theta (-E)\frac{\mathcal{T}^{3}}{8}\sum _{l=0}^{L-1}\frac{\sin [2rE\sin( \pi l/L) ]}{2r\sin( \pi l/L)\left[ 4r^{2}\sin^{2}( \pi l/L) +\mathcal{T}^{2}\right]},
\end{equation}

with $\displaystyle L=N,M$ denoting the topological charges of individual spacetimes. The interference term is

\begin{align}
\nonumber {\cal L}_{NM}(E) = & \frac{\mathcal{T}^{2}}{16} e^{-|E |\mathcal{T}}\sum _{n=0}^{N-1}\sum _{m=0}^{M-1}\frac{1}{\sqrt{NM}}\frac{1}{4r^{2}\sin^{2}[ \pi ( n/N-m/M)] +\mathcal{T}^{2}}\\\label{definition of LNM}
 & -\Theta (-E)\frac{\mathcal{T}^{3}}{8}\sum _{n=0}^{N-1}\sum _{m=0}^{M-1}\frac{1}{\sqrt{NM}}\frac{\sin[ 2rE\,\sin[ \pi ( n/N-m/M)]]}{2r\sin[ \pi ( n/N-m/M)]\left[ 4r^{2}\sin^{2}[ \pi ( n/N-m/M)] +\mathcal{T}^{2}\right]}.
\end{align}
    
\end{widetext}

These response functions possess a contribution that depends on the Heaviside function, $\Theta$, which contributes only for spontaneous decaying processes with $E<0$. For the case we are analyzing, the excitation of the UDW detector from its ground state, we need to evaluate for $E=\Delta\Omega>0$. So only the first terms in the above equation contribute for our analysis. Even more, the terms for $E>0$ posses an exponential decaying over time, which are not null only for small $E\,{\cal T}<1$. This means that a virtual excitation from the vacuum could be accomplished only for small energy gaps and short interaction times.   This behavior has been previously found in the literature  \cite{Svaiter_1994,PCWDavies_1988,Louko_2006,Juárez-Aubry_2014}.

The joint excitation and postselection response coefficient is 
\begin{multline}
\mathscr{P}(\Delta\Omega) =  \cos^{4}( \theta /2) {\cal F}_{N}( \Delta \Omega )
\\ +\sin^{4}( \theta /2) {\cal F}_{M}(\Delta \Omega )
+\frac{1}{2}\sin^{2}( \theta )\cos\!\big(\Delta\omega\,\mathcal{T}\big)\, {\cal L}_{NM}(\Delta \Omega ) 
\end{multline}
and more formally the conditional probability will be
\begin{align}
\nonumber\mathcal{P}_{e} = & \frac{\lambda ^{2}}{1-\sin^{2} \theta \ \sin^{2}\!\big(\Delta\omega\,\mathcal{T}/2\big)}\mathscr{P}(\Delta\Omega)
\end{align}

For the mixed state, we have 
\begin{equation}
\mathscr{P}_{\mathrm{cl}}(\Delta\Omega)
 = \cos^{2}\!\frac{\theta}{2}\,
   {\cal F}_{N}(\Delta\Omega)
 + \sin^{2}\!\frac{\theta}{2}\,
   {\cal F}_{M}(\Delta\Omega).
\end{equation}
which coincides with Eq.~\eqref{eq:uncond-prob} in Sec.~III when
$|c_1|^2=\cos^2(\theta/2)$ and $|c_2|^2=\sin^2(\theta/2)$.


\section{Results for Transition probabilities}

In this Section, we analyze the transition probabilities of the UDW detector in the superposed   cosmic string spacetime: 
  the transition probability $\mathscr{P}$ for the superposed state, the normalized probability ${\cal P}_e$ and the probability $\mathscr{P}_{\mathrm{cl}}$ corresponding to the mixed state. Below we show how these probabilities depend on the interaction time ${\cal T}$, the energy gap $\Delta\Omega$, the distance $r$ of the detector to the cosmic string, and the topological charges $N$ and $M$.
  All excitation probabilities reported here are computed at leading nontrivial order and therefore scale as $\lambda^{2}$. The figures show the joint postselected coefficient $\mathscr P$ and the unconditional mixed-state coefficient $\mathscr P_{\rm cl}$. Their leading probabilities are
$\lambda^2\mathscr P$ and $\lambda^2\mathscr P_{\rm cl}$. The conditional probability is ${\cal P}_e=\lambda^2\mathscr P/q_0$ where the postselection expansion is valid. 
  
The perturbative calculation requires the relevant
$\lambda^2{\cal F}_N$, $\lambda^2{\cal F}_M$, and transition
probabilities to be small. Conditional probabilities additionally
require the interaction correction to the postselection success
probability to be small compared with its zeroth-order value.

\subsection{Images of transition probability vs energy gap $\Delta\Omega$}

\begin{widetext}

\begin{figure}[hbt!]

\begin{subfigure}{0.475\linewidth}
5a) Superposed state probability $\mathscr{P}(\Delta\Omega)$
     \includegraphics[width=\linewidth]{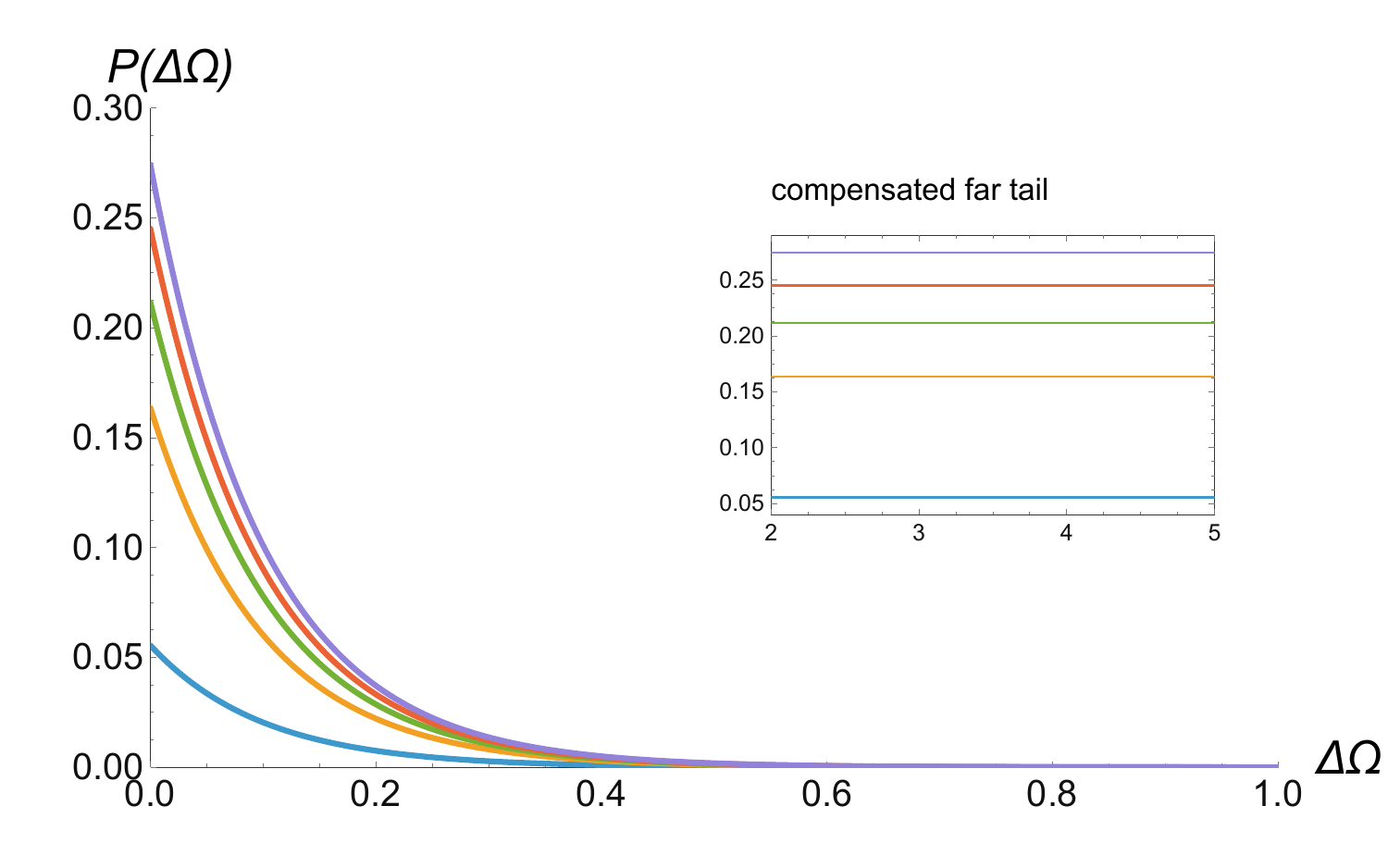}
          
          \label{PE_vs_omega_T10_r1_theta_pihalf_N4}
\end{subfigure}\hfill
\begin{subfigure}{0.475\linewidth}
5b) Mixed state probability $\mathscr{P}_{\mathrm{cl}}(\Delta\Omega)$
     \includegraphics[width=\linewidth]{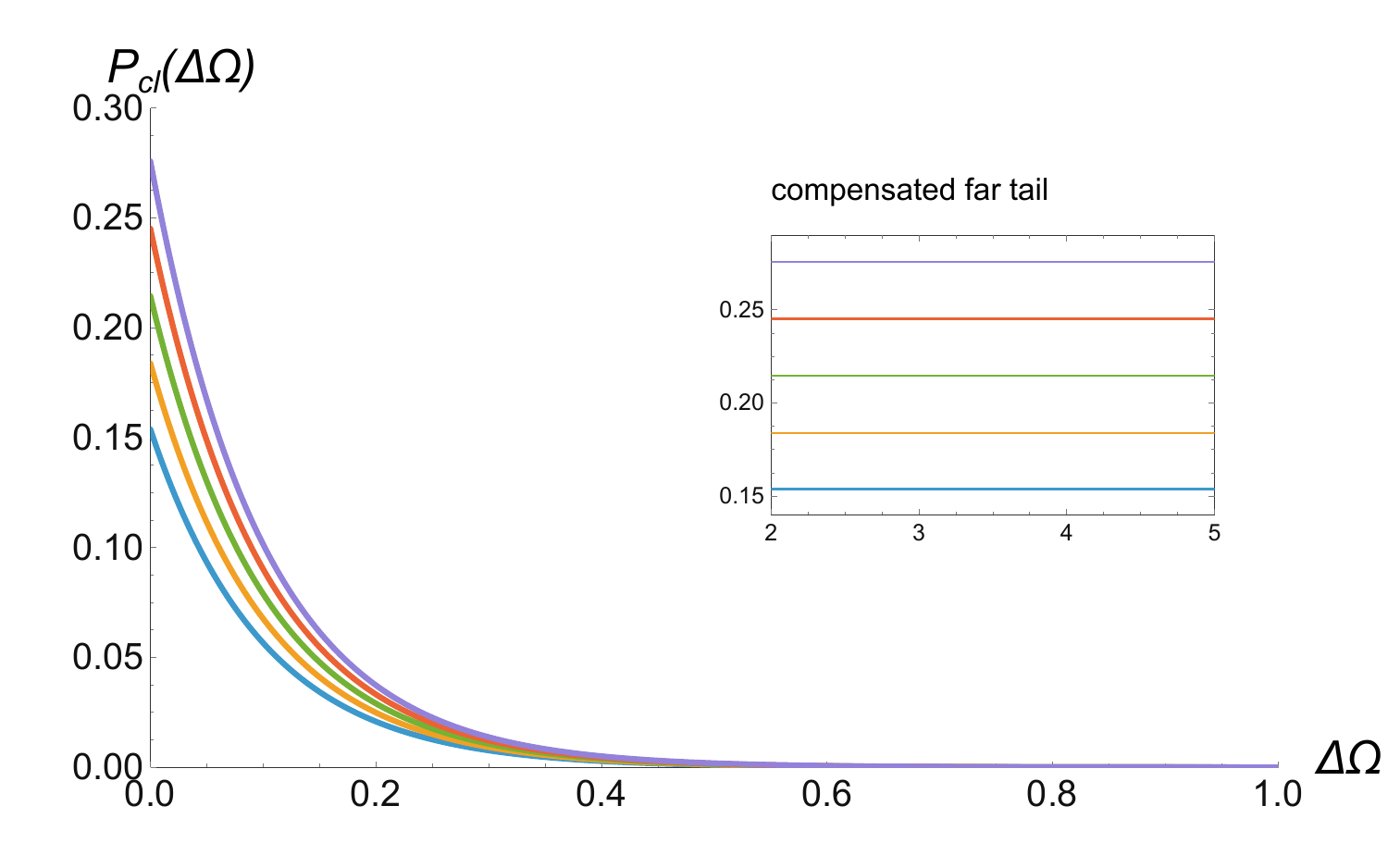}
         
          \label{ClassicPE_vs_omega_T10_r1_theta_pihalf_N4}
\end{subfigure}\hfill
\medskip 
\begin{subfigure}{0.475\linewidth}
5c) Interference deficit $-\Delta\mathscr{P}_{\mathrm{sup}}(\Delta\Omega)$
     \includegraphics[width=\linewidth]{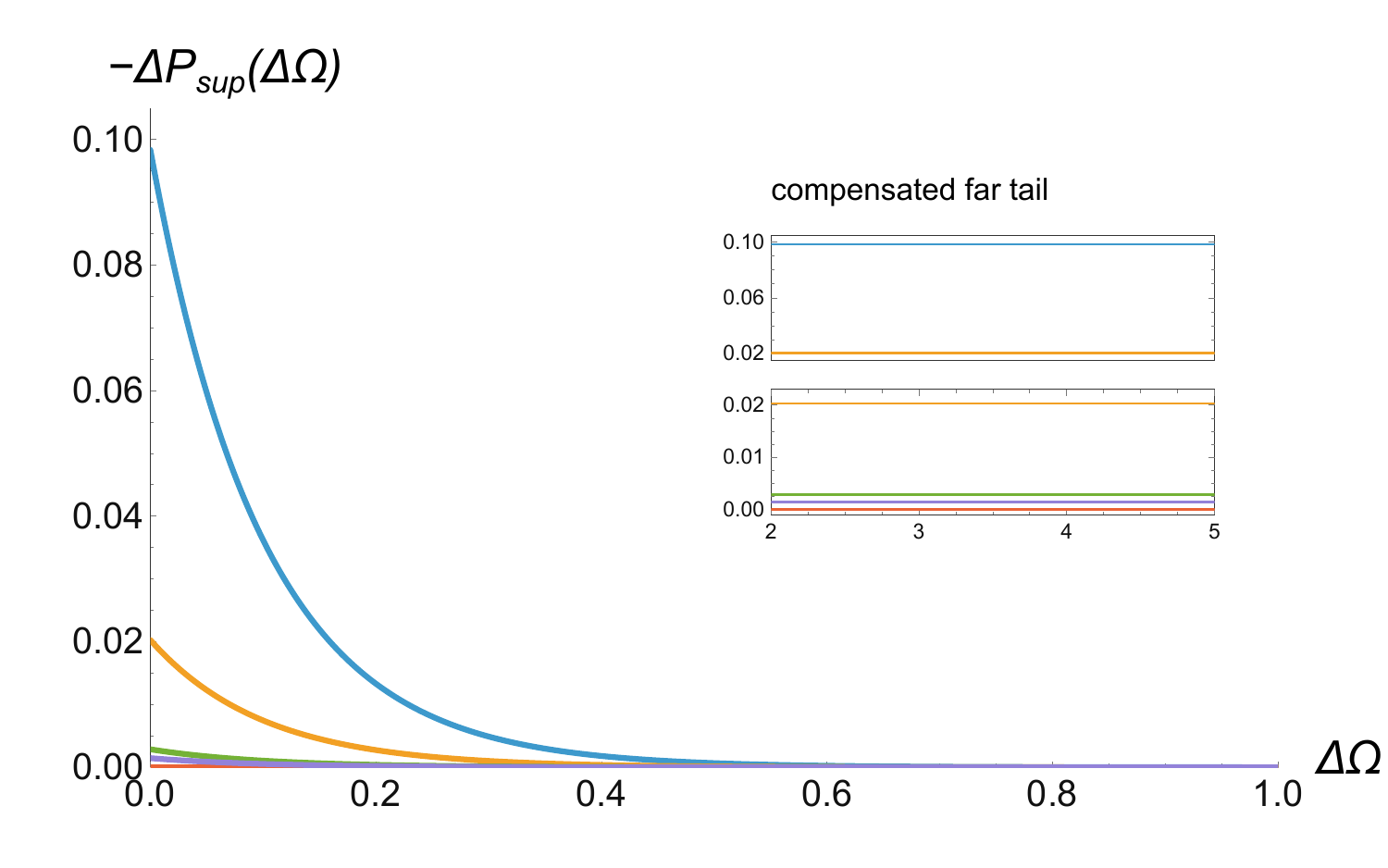}
        
          \label{PEn_vs_omega_T10_r1_theta_pihalf_N4}
\end{subfigure}\hfill
\begin{subfigure}{0.475\linewidth}
5d) Ratio: superposed and mixed states $\mathscr{P}(\Delta\Omega)/\mathscr{P}_{\mathrm{cl}}(\Delta\Omega)$
     \includegraphics[width=\linewidth]{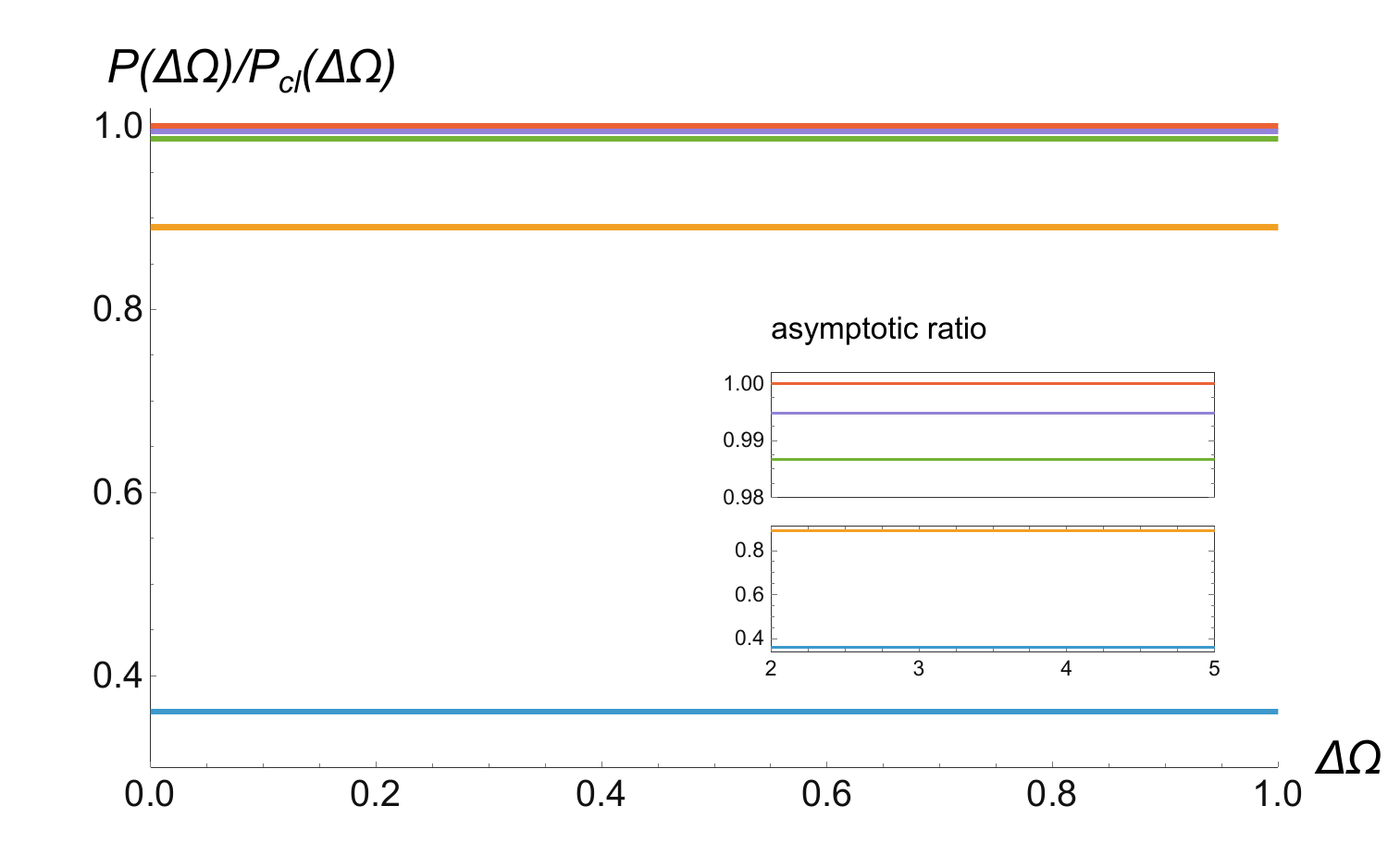}
          
          \label{PEQPEC_vs_omega_T10_r1_theta_pi_half_N4}
\end{subfigure}\hfill
\caption{Dependence of the excitation probabilities on the UDW energy gap, evaluated for $\mathcal{T}=10$, $r=1$, $\theta=\pi/2$, and $N=4$. The common colour key for panels (a)--(d) is \protect\mnlegend.
}
\label{PE_vs_omega}
\end{figure}
\end{widetext}

  We begin by considering how the different transition probabilities depend on the detector energy gap $\Delta\Omega>0$.
Our results are shown in figure~\ref{PE_vs_omega}.

For large gap the excitation transition probability decreases rapidly;  vacuum fluctuations hardly modify the state of a qubit initially in the ground state. For small gap, the detector
can become excited even if it is in an inertial state of motion
due to the topological character of the string \cite{Svaiter_1994,PCWDavies_1988}.   The transition probability increases with increasing topological charge.  

Figure~\ref{PE_vs_omega} shows how the excitation probability decays with increasing detector gap $\Delta\Omega>0$. The positive-gap responses share the exact switching factor $e^{-\Delta\Omega\mathcal{T}}$, so their unscaled large-gap curves all vanish exponentially and cannot display their limiting ordering once they collapse onto the horizontal axis. The insets in panels (a)--(c) therefore show the compensated quantities $e^{\Delta\Omega\mathcal{T}}\mathscr{P}$, $e^{\Delta\Omega\mathcal{T}}\mathscr{P}_{\mathrm{cl}}$, and $e^{\Delta\Omega\mathcal{T}}(\mathscr{P}_{\mathrm{cl}}-\mathscr{P})$ over $2\leq\Delta\Omega\leq5$, corresponding to $20\leq\Delta\Omega\mathcal{T}\leq50$. Their horizontal plateaus are the asymptotic coefficients. Panels (a) and (b) retain the order $M_1<M_2<M_3<M_4<M_5$, whereas the deficit in panel (c) obeys $M_4<M_5<M_3<M_2<M_1$; 
no positive-gap ordering switch occurs. This ordering highlights the fact that the superposed case increasingly differs most from the mixed case as the difference in topological charge increases.
Panel (d) compares the coherent superposition with the incoherent mixture. Its inset shows the same far-gap interval and confirms the constant asymptotic ratio ordering $M_1<M_2<M_3<M_5<M_4$, with $M_4=N$ exactly unity. All main-panel curves are displayed through $\Delta\Omega=0$: the intercepts in panels (a) and (b) are finite and generally nonzero, while the $M_4=N$ curve in panel (c) vanishes identically because there is no branch mismatch.

\subsection{Images of transition probability vs interaction time $\mathcal{T}$}

The dependence on the interaction time $\mathcal{T}$ is shown in figure~\ref{PE_vs_T}. At fixed detector gap $\Delta\Omega=0.1$, the excitation probabilities reach their maxima at short but nonzero interaction times and are exponentially suppressed for $\Delta\Omega\,\mathcal{T}\gg1$ \cite{Louko_2006,Juárez-Aubry_2014,Enrike}.For fixed $r>0$, the peak-one Lorentzian gives ${\cal F}_L\to1/16$ as $\mathcal T\to0^+$.
This finite limit reflects the ultraviolet scaling of the
pointlike detector.

The joint and unconditional responses are compared in figure~\ref{PE_vs_T}(c) and (d). Panel (c) shows the nonnegative deficit $-\Delta\mathscr{P}_{\mathrm{sup}}=\mathscr{P}_{\mathrm{cl}}-\mathscr{P}$, while panel (d) compares the coherent superposition with the mixed (unconditional) case. The oscillation period and amplitude depend on the topological mismatch through the relative branch phase. The bound $\mathscr{P}/\mathscr{P}_{\mathrm{cl}}\leq1$ does not by itself establish destructive interference. When $M_4=N$, the two branches coincide: the deficit vanishes and the ratio remains exactly unity.

In figure~\ref{PE_vs_T_vs_W} we plot the coherent and mixed responses as functions of both $\mathcal{T}$ and $\Delta\Omega$. For $N=3$ and $M=2$, the relative branch frequency is $|\Delta\omega|=1/24$ in the units used here, so the interference period is $2\pi/|\Delta\omega|=48\pi\simeq151$. We therefore display $0\leq\mathcal{T}\leq320$, which resolves two oscillations, and focus on $0\leq\Delta\Omega\leq0.02$, where they are not erased by the factor $e^{-\Delta\Omega\mathcal{T}}$. The quantum surface exhibits the resulting alternating enhancement and suppression at low gap, whereas the surface for the mixed case is smooth because it contains no relative-phase interference term.

\begin{widetext}
    
\begin{figure}[hbt!]
\begin{subfigure}{0.475\linewidth}
6a) Superposed state probability $\mathscr{P}(\mathcal{T})$
\centering
     \includegraphics[width=\linewidth]{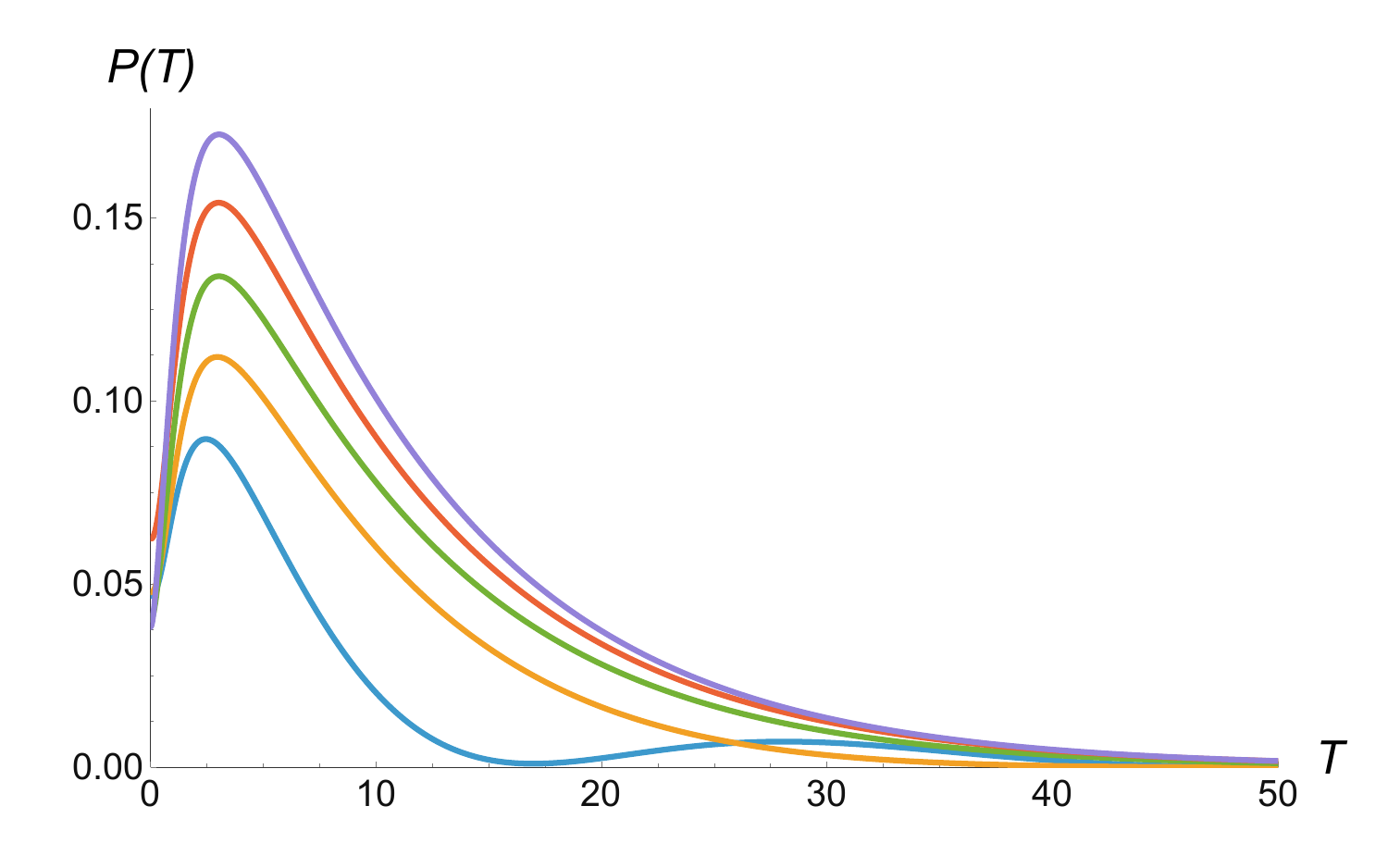}
          \label{PE_vs_T_omega1_r1_theta_pihalf_N4}
\end{subfigure}\hfill
\begin{subfigure}{0.475\linewidth}
6b) Mixed state probability $\mathscr{P}_{\mathrm{cl}}(\mathcal{T})$
\centering
     \includegraphics[width=\linewidth]{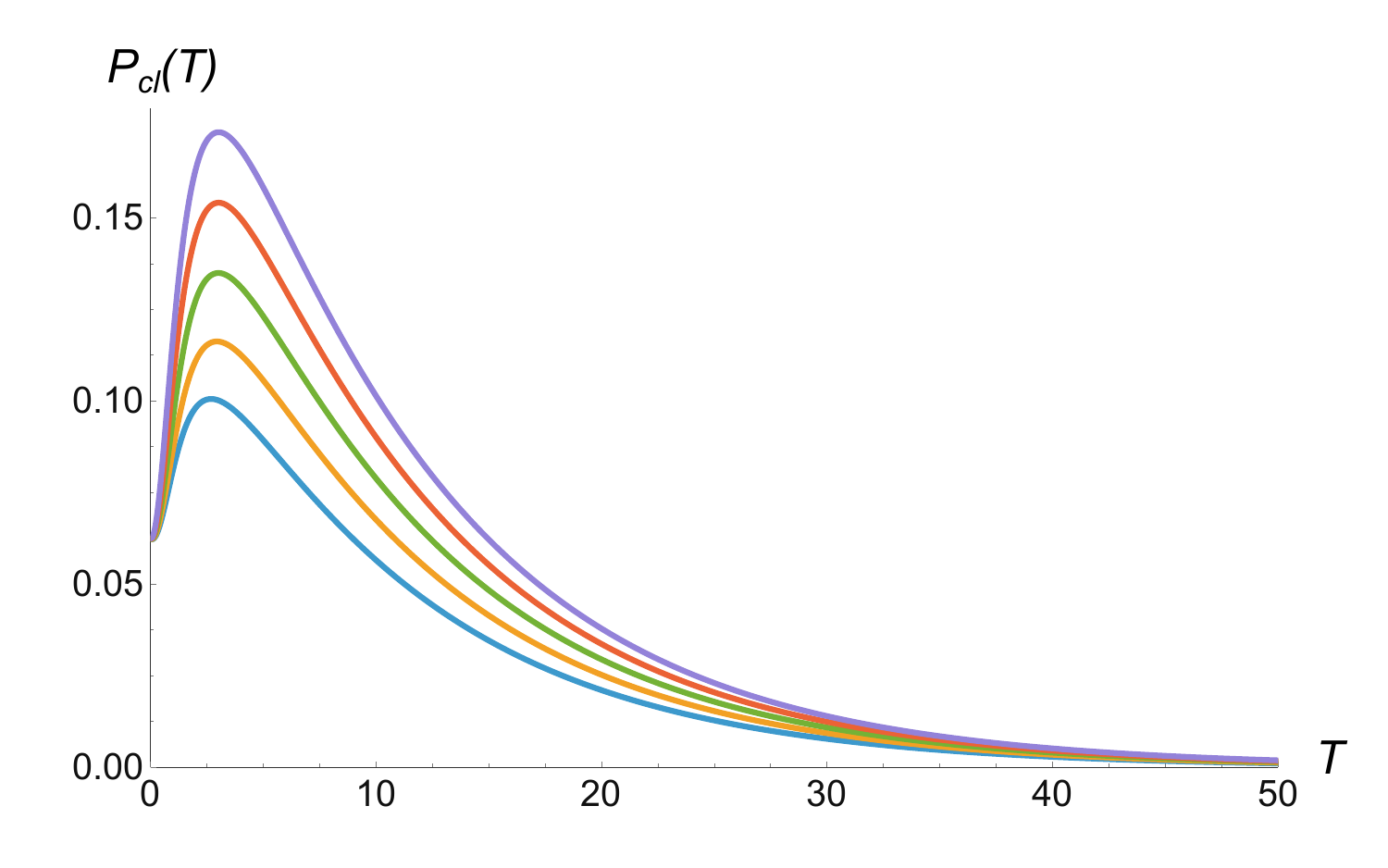}
          \label{ClassicalPE_vs_T_omega1_r1_theta_pihalf_N4}
\end{subfigure}\hfill
\medskip
\begin{subfigure}{0.475\linewidth}
6c) Interference deficit $-\Delta\mathscr{P}_{\mathrm{sup}}(\mathcal{T})$
\centering
     \includegraphics[width=\linewidth]{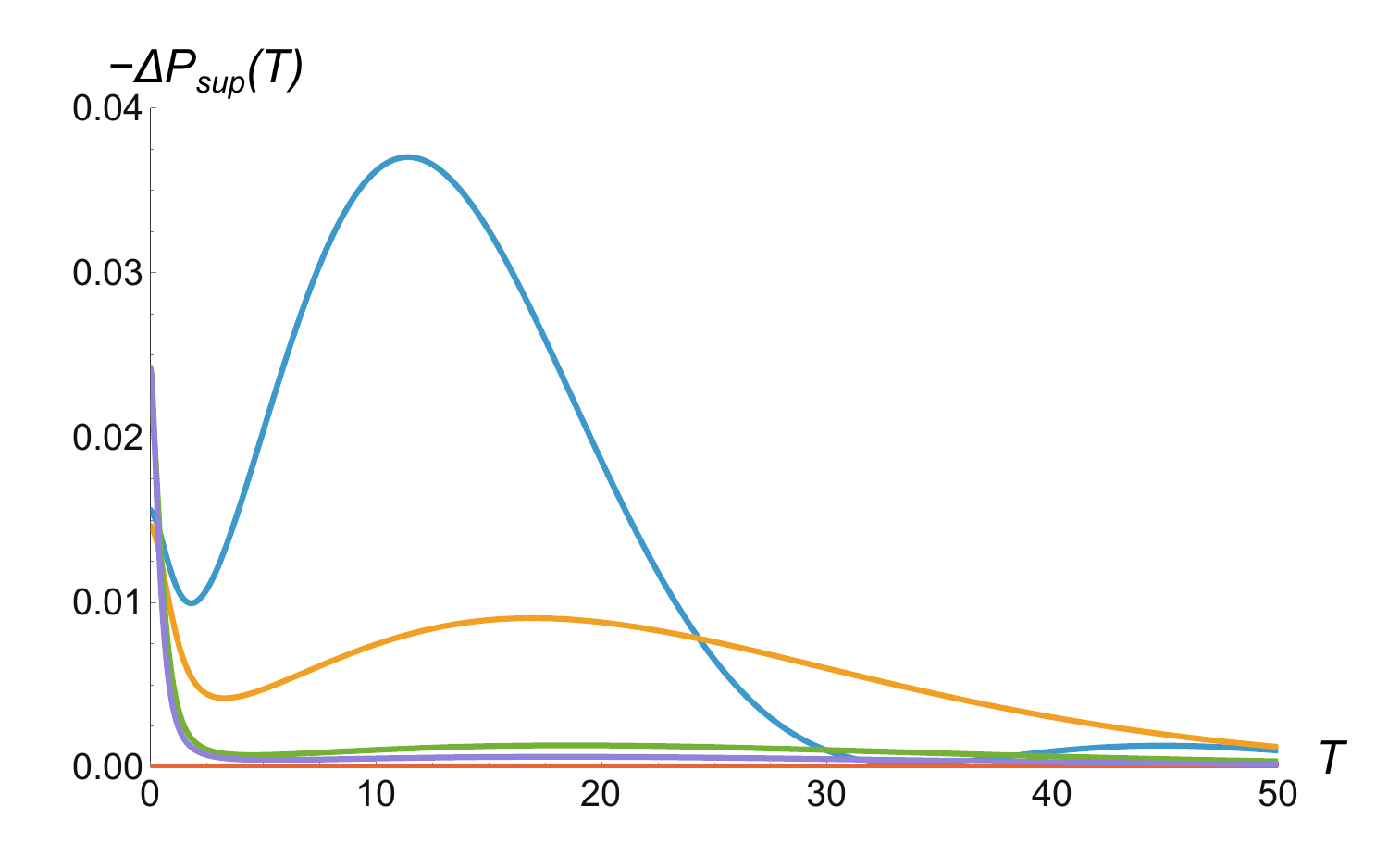}
       
          \label{PEn_vs_T_omega1_r1_theta_pihalf_N4}
\end{subfigure}\hfill
\begin{subfigure}{0.475\linewidth}
6d) Ratio $\mathscr{P}(\mathcal{T})/\mathscr{P}_{\mathrm{cl}}(\mathcal{T})$
\centering
     \includegraphics[width=\linewidth]{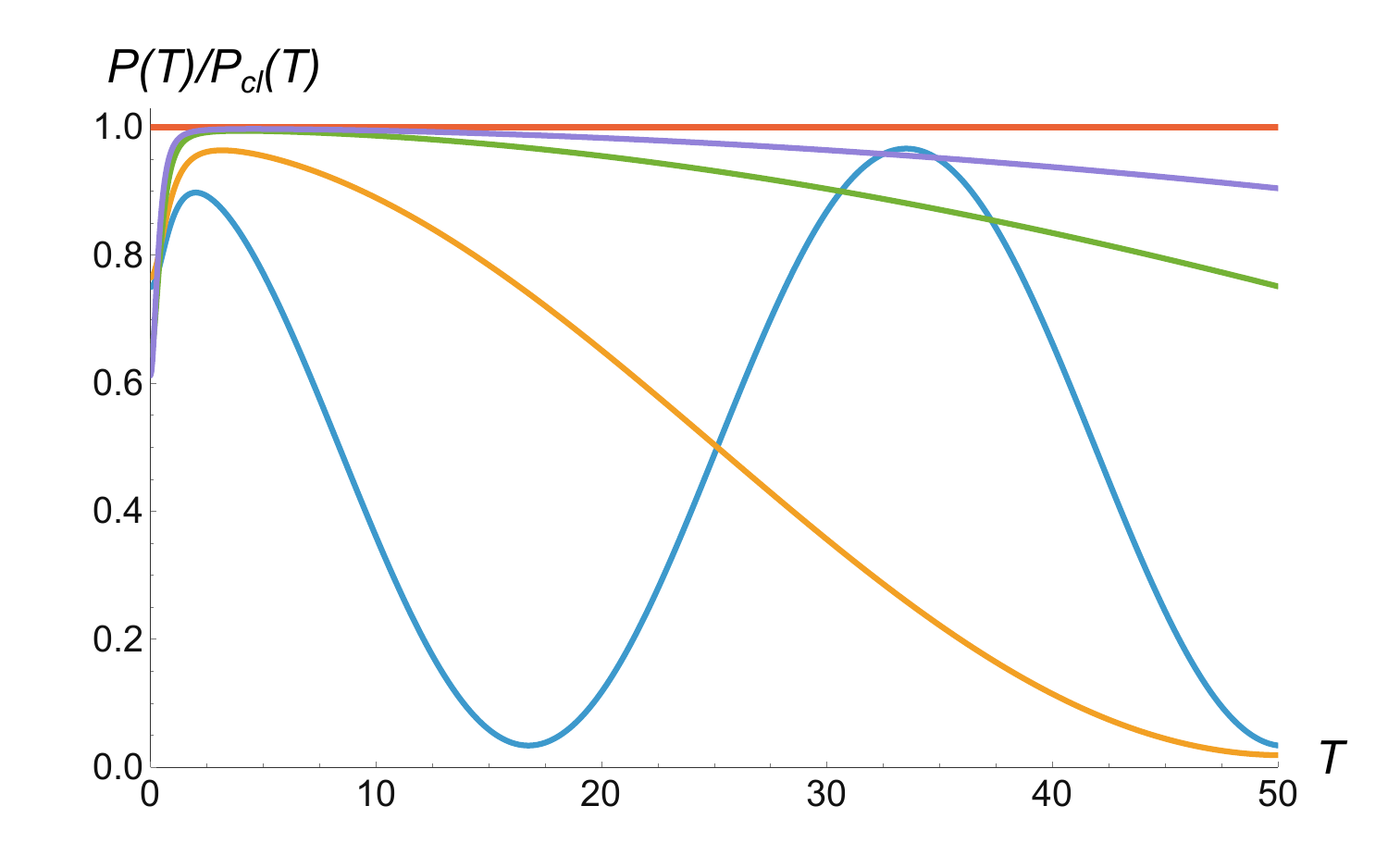}
          
          \label{PEQPEC_vs_T_omega1_r1_theta_pi_half_N4}
\end{subfigure}\hfill

\caption{Interaction-time dependence of the detector response, evaluated for $\Delta\Omega=0.1$, $r=1$, $\theta=\pi/2$, and $N=4$. The common colour key for panels (a)--(d) is \protect\mnlegend. 
}
\label{PE_vs_T}
\end{figure}

\begin{figure}[hbt!]
\begin{subfigure}{0.475\linewidth}
(a) Superposed state probability $\mathscr{P}(\mathcal{T},\Delta\Omega)$
\centering
     \includegraphics[width=\linewidth]{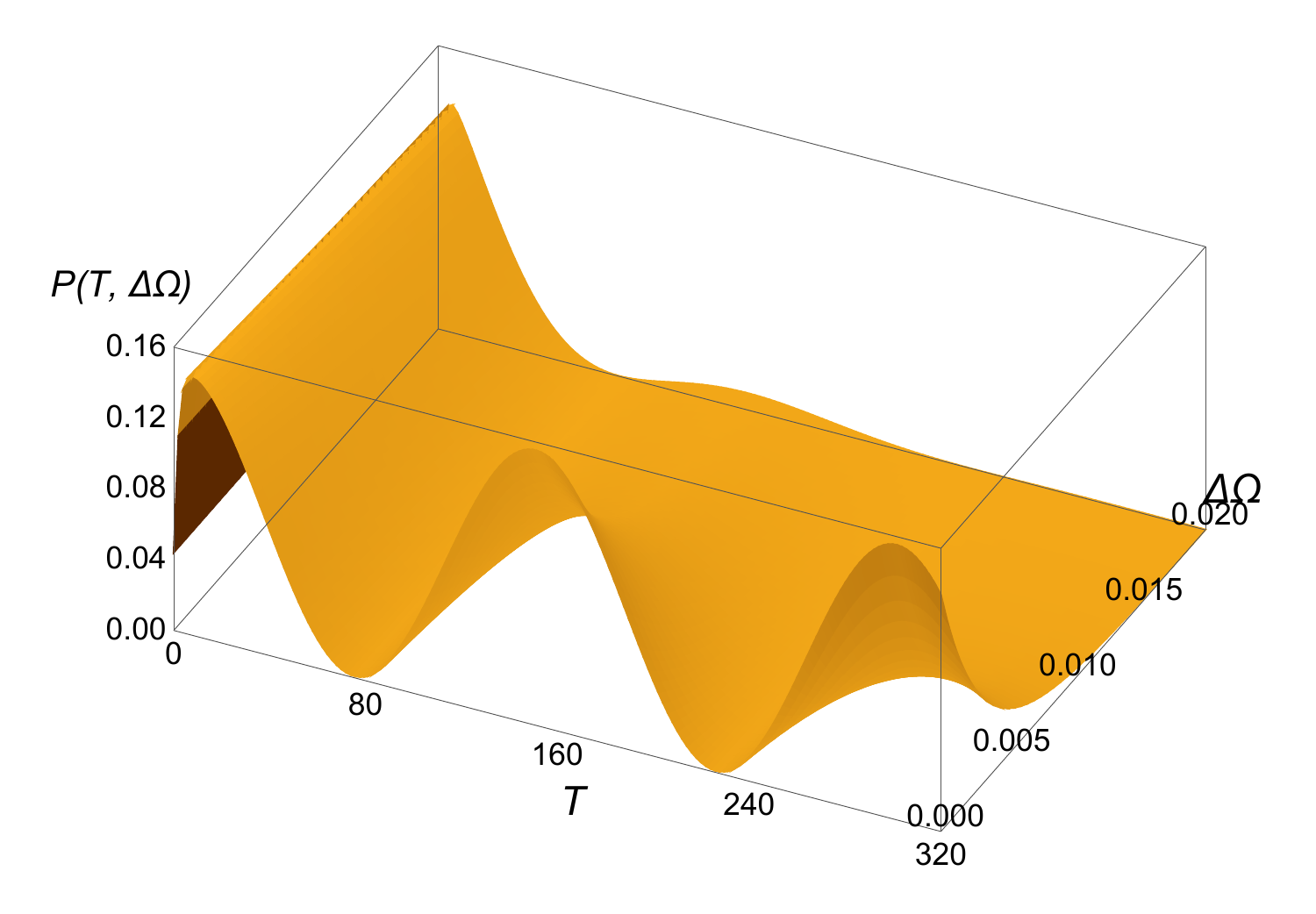}
          \label{PE_T_vs_omega_N3_M2.pdf}
\end{subfigure}\hfill
\begin{subfigure}{0.475\linewidth}
(b) Mixed state probability $\mathscr{P}_{\mathrm{cl}}(\mathcal{T},\Delta\Omega)$
\centering
     \includegraphics[width=\linewidth]{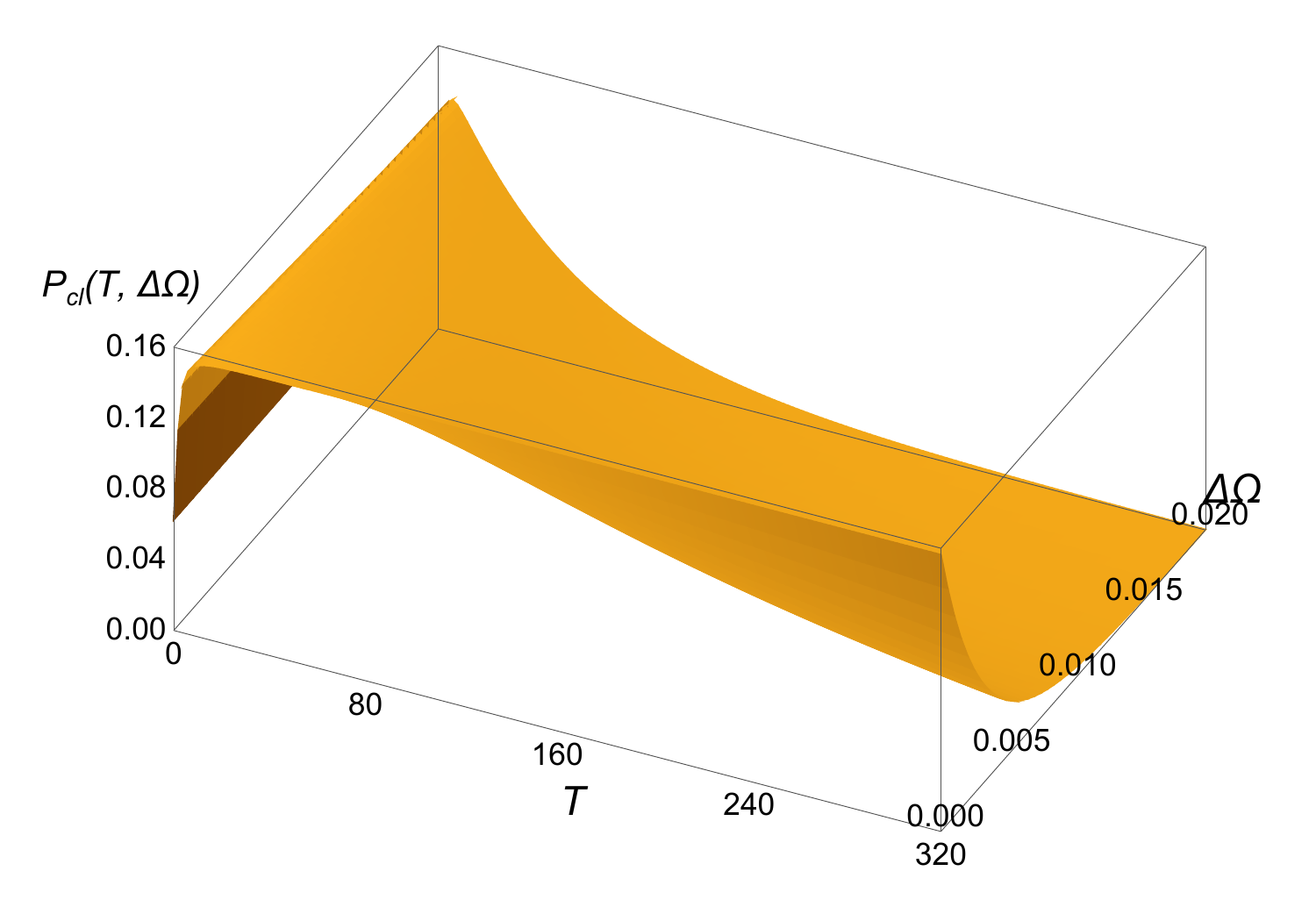}
          \label{ClassicalPE_T_vs_omega_N3_M2.pdf}
\end{subfigure}\hfill
\caption{Quantum and mixed state excitation probabilities as functions of interaction time and detector energy gap, evaluated for $r=1$, $\theta=\pi/2$, $N=3$, and $M=2$. The displayed ranges are $0\leq\mathcal{T}\leq320$ and $0\leq\Delta\Omega\leq0.02$; the enlarged low-gap window makes the phase-driven oscillations in panel (a) visible, while panel (b) provides the nonoscillatory mixed-state comparison.}
\label{PE_vs_T_vs_W}
\end{figure}

\begin{figure}[hbt!]
\begin{subfigure}{0.475\linewidth}
(a) Superposed state probability $\mathscr{P}(r/\mathcal{T})$
\centering
     \includegraphics[width=\linewidth]{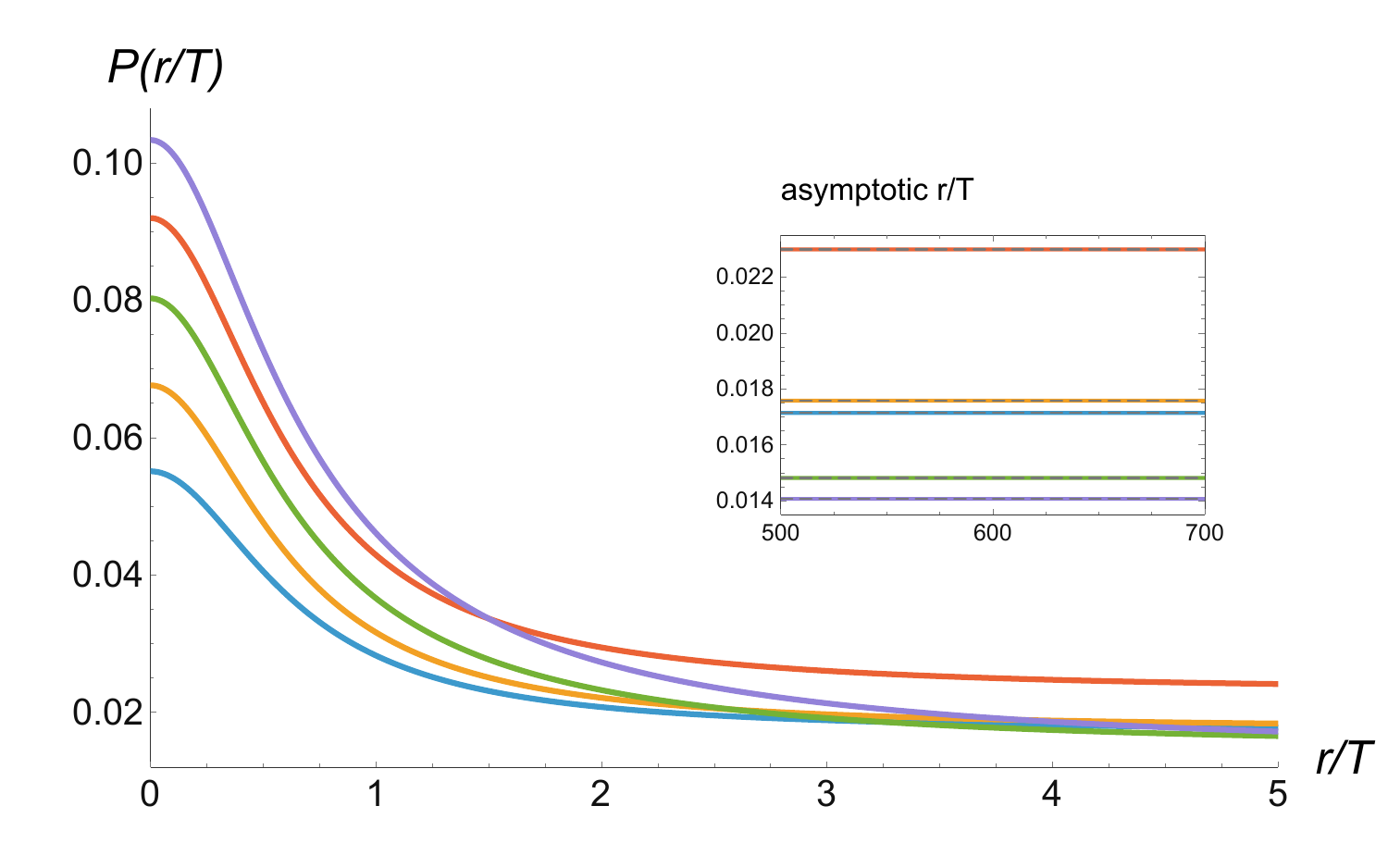}
          \label{PE_vs_r_T1_omega1_theta_pihalf_N4}
\end{subfigure}\hfill
\begin{subfigure}{0.475\linewidth}
(b) Mixed state probability $\mathscr{P}_{\mathrm{cl}}(r/\mathcal{T})$
\centering
     \includegraphics[width=\linewidth]{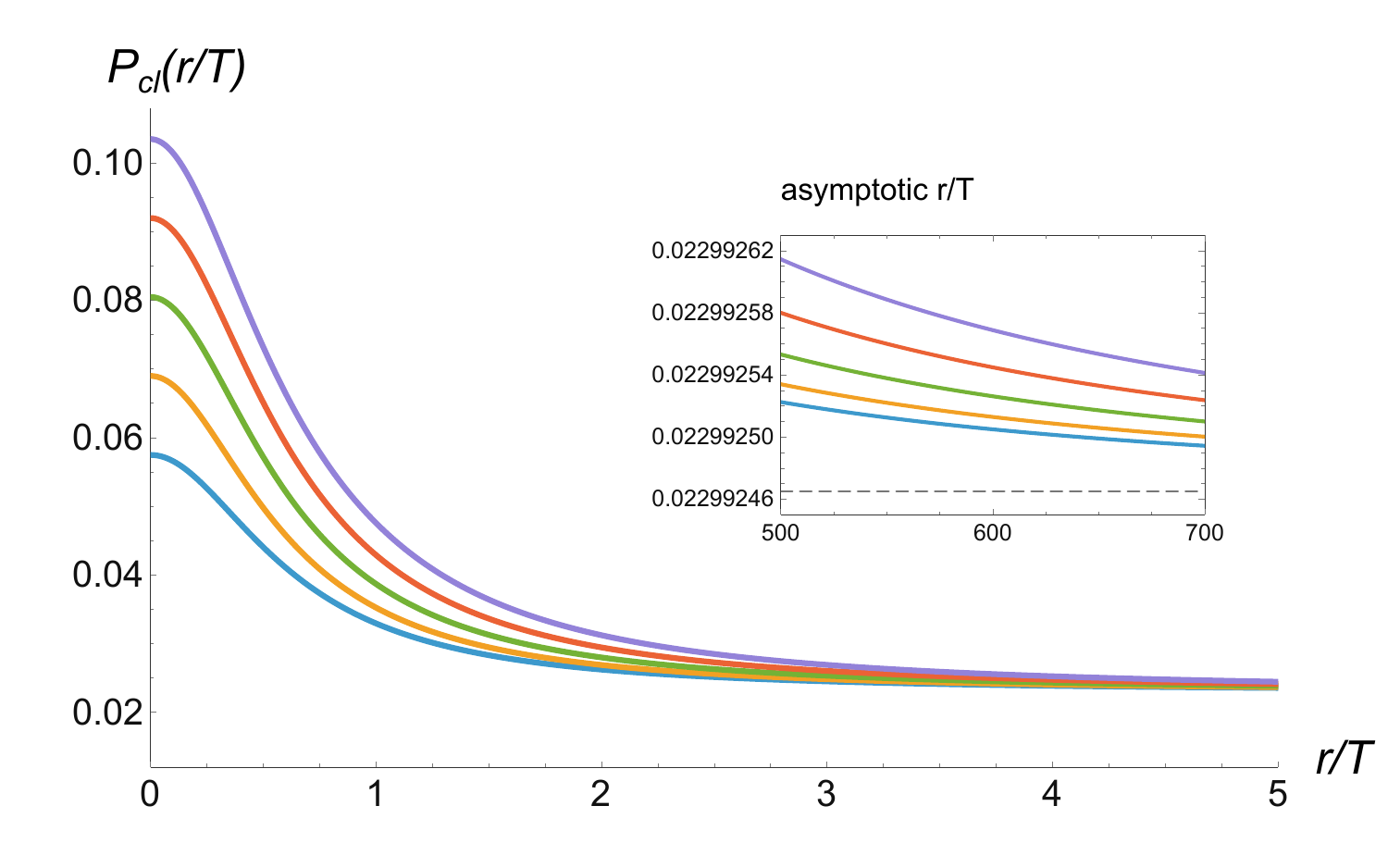}
          \label{ClassicalPE_vs_r_T1_omega1_theta_pihalf_N4}
\end{subfigure}\hfill
\medskip
\begin{subfigure}{0.475\linewidth}
(c) Interference deficit $\mathscr{P}_{\mathrm{cl}}-\mathscr{P}$
\centering
     \includegraphics[width=\linewidth]{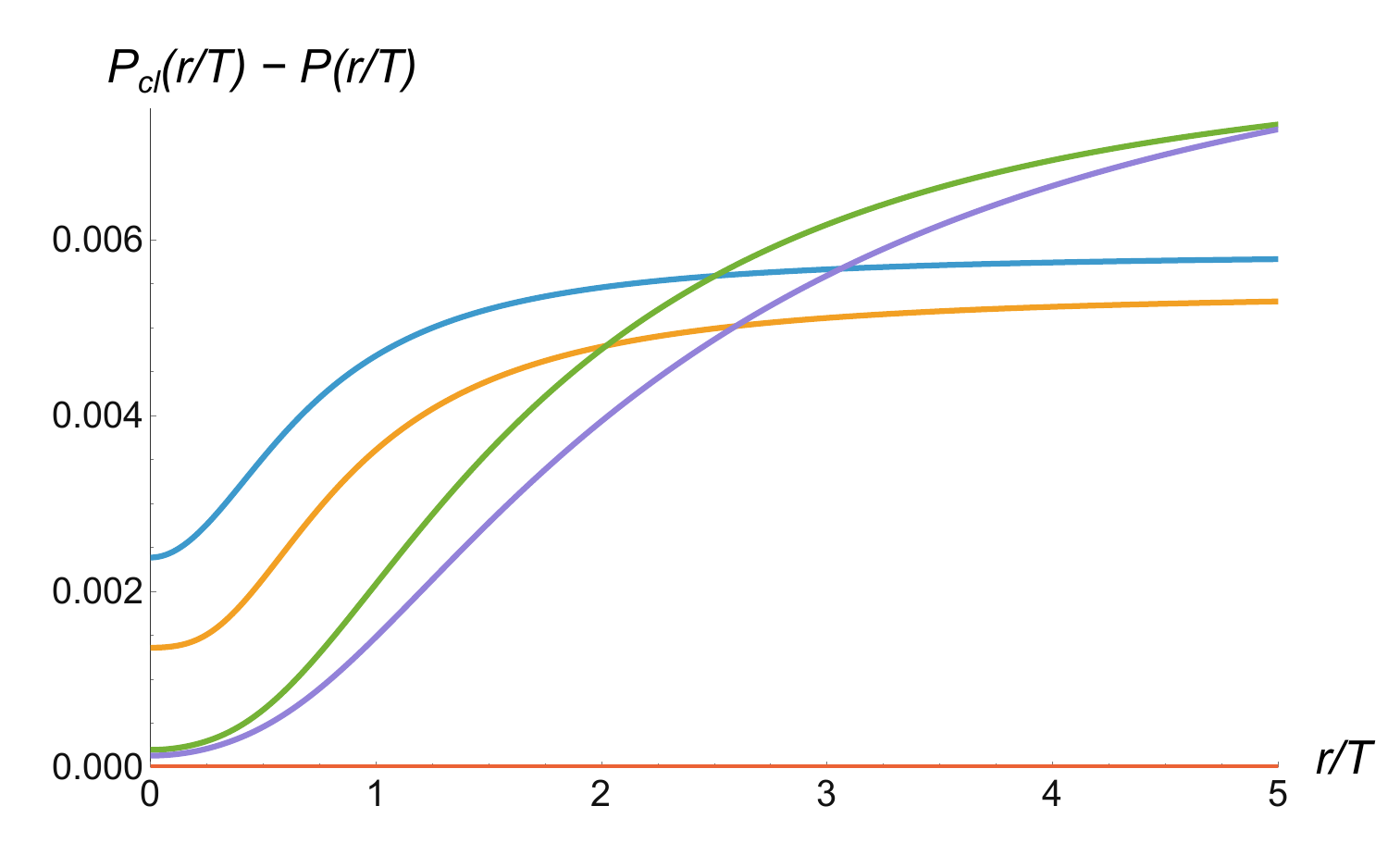}
          \label{PEn_vs_r_T1_omega1_theta_pihalf_N4}
\end{subfigure}\hfill
\begin{subfigure}{0.475\linewidth}
(d) Ratio $\mathscr{P}(r/\mathcal{T})/\mathscr{P}_{\mathrm{cl}}(r/\mathcal{T})$
\centering
     \includegraphics[width=\linewidth]{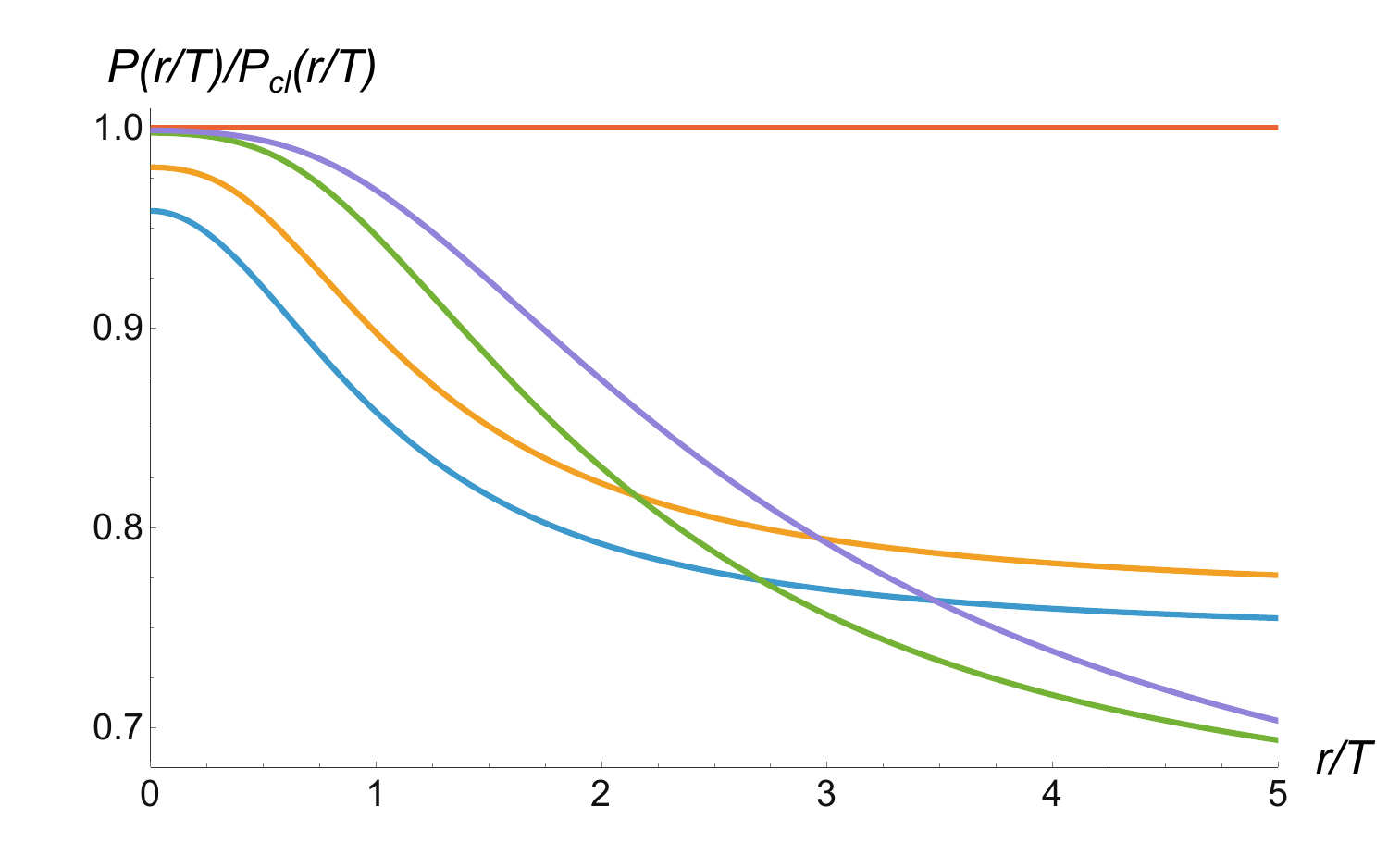}
          \label{PEQPEC_vs_r_T1_omega1_theta_pi_half_N4}
\end{subfigure}\hfill
\caption{Dependence on the dimensionless detector--string distance $r/\mathcal{T}$, evaluated for $\mathcal{T}=1$, $\Delta\Omega=1$, $\theta=\pi/2$, and $N=4$. The common colour key for panels (a)--(d) is \protect\mnlegend. The insets in panels (a) and (b) show the asymptotic interval $500\leq r/\mathcal{T}\leq700$; the grey dashed lines mark the corresponding $r/\mathcal{T}\to\infty$ limits. Panel (a) resolves the distinct quantum plateaus in the order $M_5<M_3<M_1<M_2<M_4$, whereas panel (b) shows all mixed state curves converging to their common limit. Panel (c) plots the nonnegative deficit $\mathscr{P}_{\mathrm{cl}}-\mathscr{P}$ directly; its $M_4=N$ curve is identically zero and lies on the horizontal axis.}
\label{PE_vs_r}
\end{figure}

\begin{figure}[hbt!]
\begin{subfigure}{0.475\linewidth}
(a) Superposed state probability $\mathscr{P}(\theta)$
\centering
     \includegraphics[width=\linewidth]{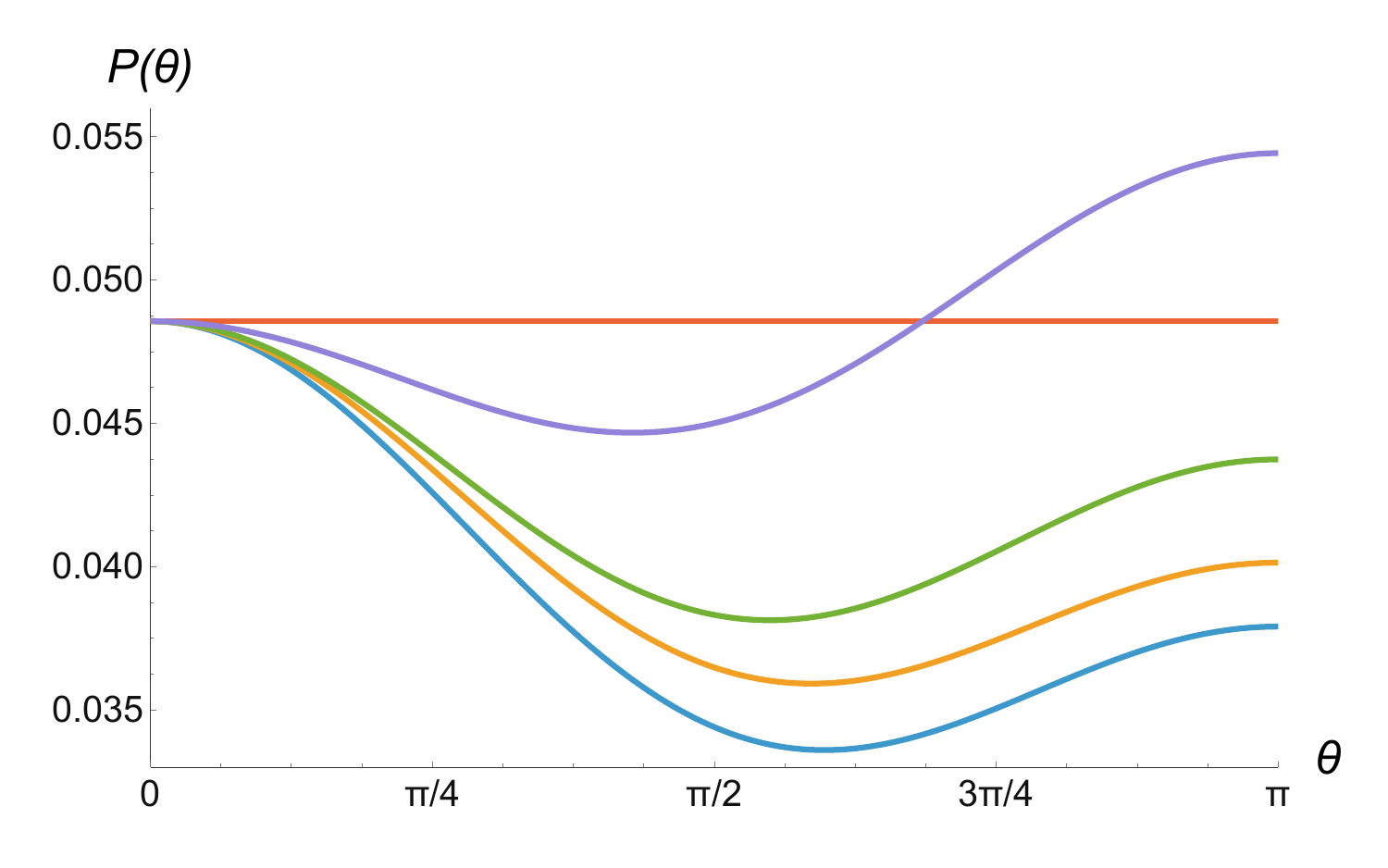}
          \label{PE_vs_theta_T1_omega1_r1_N4}
\end{subfigure}\hfill
\begin{subfigure}{0.475\linewidth}
(b) Mixed state probability $\mathscr{P}_{\mathrm{cl}}(\theta)$
\centering
     \includegraphics[width=\linewidth]{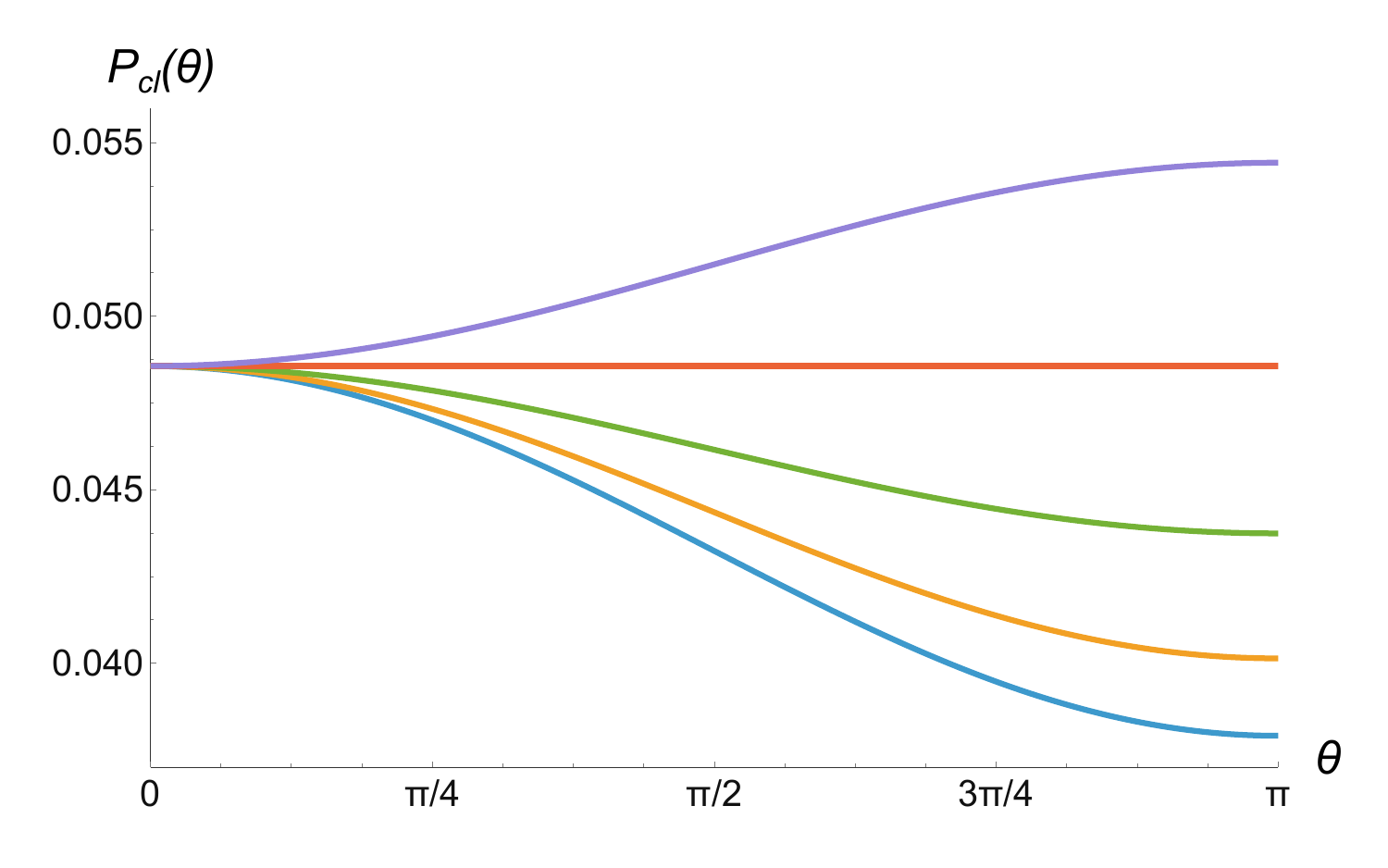}
          \label{ClassicalPE_vs_theta_T1_omega1_r1_N4}
\end{subfigure}\hfill
\medskip
\begin{subfigure}{0.475\linewidth}
(c) Interference deficit $\mathscr{P}_{\mathrm{cl}}(\theta)-\mathscr{P}(\theta)$
\centering
     \includegraphics[width=\linewidth]{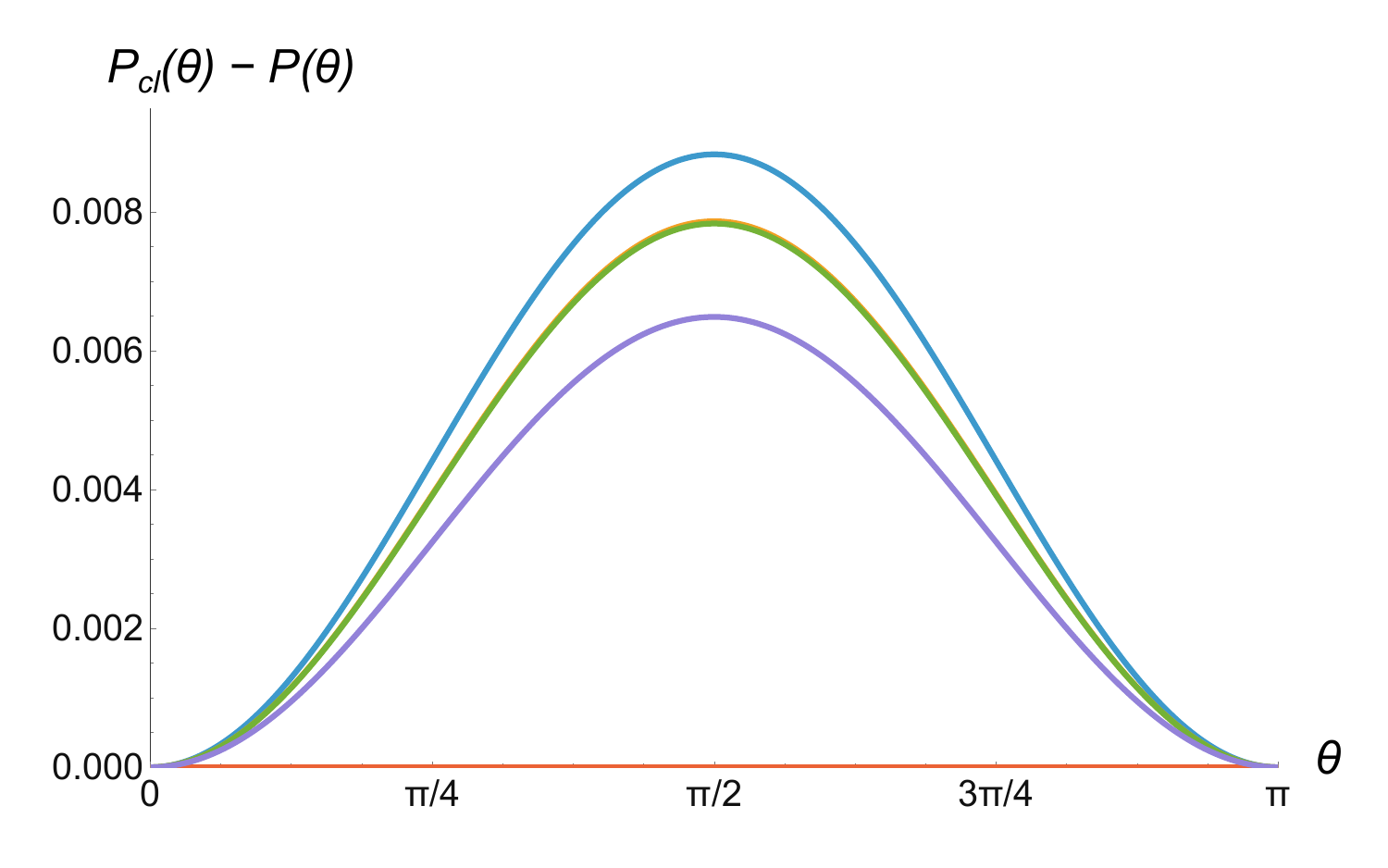}
          \label{PEn_vs_theta_T1_omega1_r1_N4}
\end{subfigure}\hfill
\begin{subfigure}{0.475\linewidth}
(d) Ratio $\mathscr{P}(\theta)/\mathscr{P}_{\mathrm{cl}}(\theta)$
\centering
     \includegraphics[width=\linewidth]{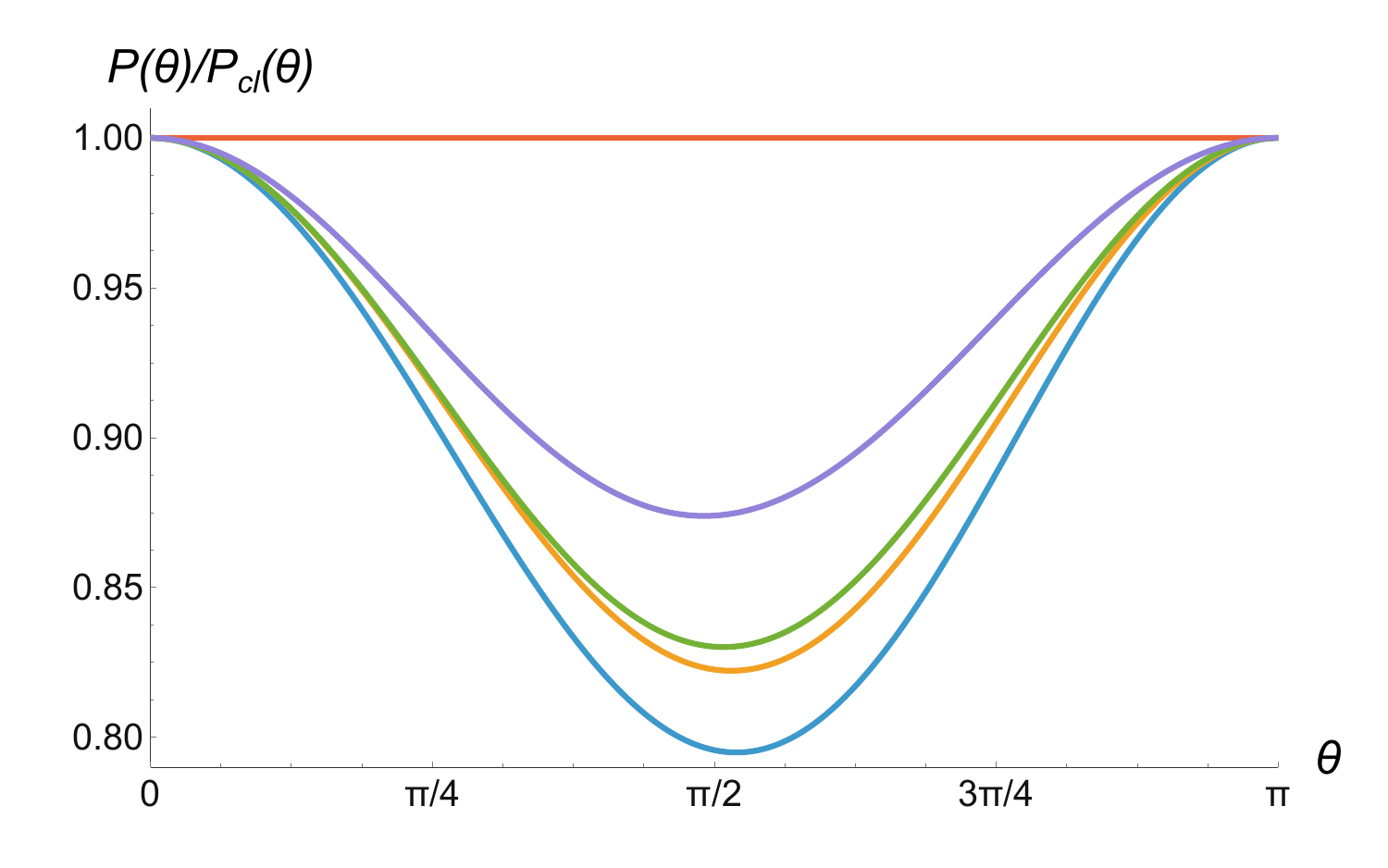}
          \label{PEQPEC_vs_theta_T05_omega1_r1_N4}
\end{subfigure}\hfill

\caption{Dependence on the superposition angle $\theta$, evaluated for $\mathcal{T}=0.5$, $\Delta\Omega=1$, $r=1$, and $N=4$. The common colour key for panels (a)--(d) is \protect\mnlegend. Note: Panel (c) plots the nonnegative interference deficit $\mathscr{P}_{\mathrm{cl}}-\mathscr{P}$ directly; its $M_4=N$ curve is identically zero and lies on the horizontal axis. At the single-branch endpoints $\theta=0,\pi$, the deficit vanishes and the ratio in panel (d) equals unity.}
\label{PE_vs_theta}
\end{figure}

\begin{figure}[ht]
\centering
     \includegraphics[width=\linewidth]{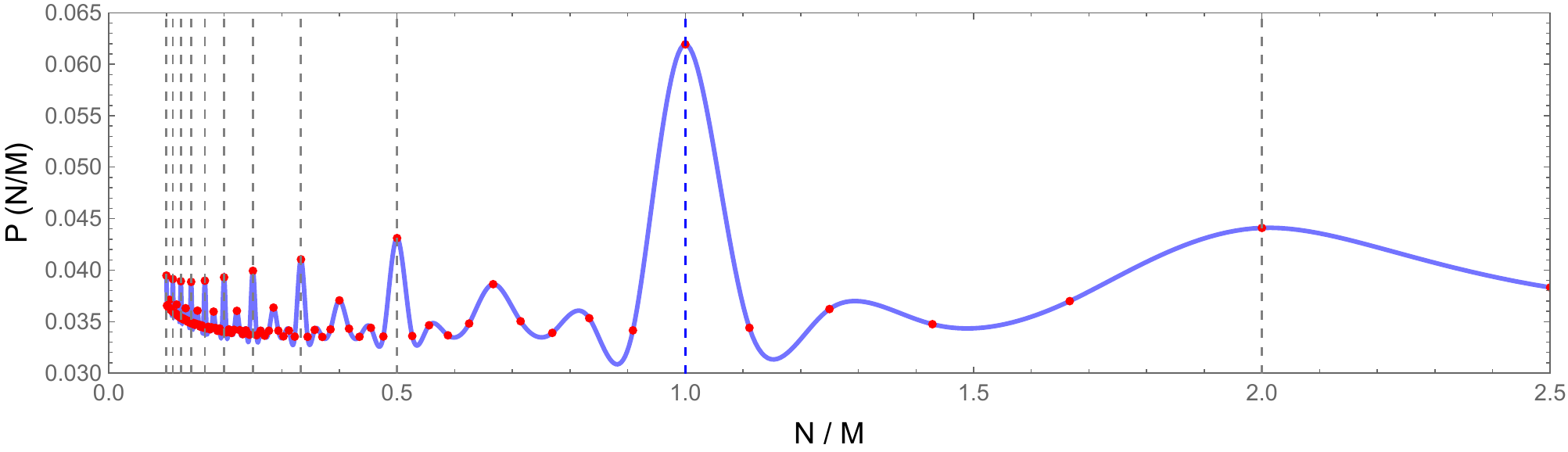}
          \caption{Joint transition response coefficient as a function of the ratio of cosmic string topological charges $N/M$. We use $\Omega=1$, $r=1$, ${\cal T}=0.01$,  $\theta=\pi/2$, $N=10$ and $M=\{1,...,100\}$. The gray dashed lines indicate secondary maxima obtained at $M=\{5,20,30,...,100\}$while the blue dashed lines indicate the principal maximum obtained at the resonance $M=N=10$ \cite{ALIEV1989142,smith1989gravitational,Volovik:1998de,Volovik:1996qw,Barcelo:2005fc}. Note that the solid blue line is an interpolation to guide the eye; there are no points in the graph apart from the red dots.}
          \label{PE_vs_gamma_T0.01_omega1_r1_theta_pihalf_N1to100_M10}         
\end{figure}

\end{widetext}

\subsection{Images of transition probability vs dimensionless cosmic-string distance $r/\mathcal{T}$}

We now consider the detector--string separation through the dimensionless ratio $r/\mathcal{T}$, which is common to both branches of the superposition. The choice $M=1$ denotes a Minkowski (no-defect) branch, although the state plotted here still contains the fixed $N=4$ branch. For each individual conical branch, the image contributions vanish as $r/\mathcal{T}\to\infty$, leaving the Minkowski response $\mathcal{F}_1=e^{-\Delta\Omega\mathcal{T}}/16$ \cite{Svaiter_1994,PCWDavies_1988}. Consequently, all curves in the  mixed case figure~\ref{PE_vs_r}(b) approach the same $M$-independent limit; its $500\leq r/\mathcal{T}\leq700$ inset shows the curves flattening onto that analytic plateau rather than merely beginning their convergence.

At the string, $r=0$, the branch response obeys $\mathcal{F}_L=L\mathcal{F}_1$ \cite{Fulling_2012,ALIEV1989142}. For the balanced mixture used here, $\mathscr{P}_{\mathrm{cl}}(0)=(N+M)\mathcal{F}_1/2$, so the near-string excitation increases monotonically with $M$ and the $M=1$ case is the smallest. The postselected quantum response in panel (a) has a different large-distance structure: its surviving cross-branch coherence produces $M$-dependent asymptotes, with the coincident-branch case $M_4=N$ recovering the Minkowski value. Panel (c) contains all five curves; the apparently missing $M_4$ curve is the zero-deficit line on the axis.

\subsection{Images of transition probability vs superposition angle $\theta$}

The angle $\theta$ is the branch-mixing (Bloch-sphere) parameter of the initial spacetime state,
\begin{equation}
|\mu\rangle=\cos(\theta/2)\,|\mu_1\rangle+e^{i\phi}\sin(\theta/2)\,|\mu_2\rangle,
\end{equation}
as shown geometrically in Fig.~\ref{fig:enter-label}.  In particular, $\theta=0$ and $\theta=\pi$ correspond to single-branch geometries ($|\mu\rangle=|\mu_1\rangle$ or $|\mu_2\rangle$), while $\theta=\pi/2$ gives a balanced superposition with $|c_1|^2=|c_2|^2=1/2$.  The joint excitation and postselection coefficient depends on $\theta$ through the branch weights and the interference amplitude:
\begin{multline}
\mathscr{P}(\Delta\Omega)
= \cos^{4}\!\frac{\theta}{2}\,{\cal F}_{N}(\Delta \Omega )
+\sin^{4}\!\frac{\theta}{2}\,{\cal F}_{M}(\Delta \Omega )\\
+\frac{1}{2}\sin^{2}\!\theta \;\cos\!\big(\Delta\omega\,\mathcal{T}\big)\,{\cal L}_{NM}(\Delta \Omega ),
\label{eq:Ptheta-discussion}
\end{multline}
whereas the  (unconditional) mixture is the convex sum
\begin{equation}
\mathscr{P}_{\mathrm{cl}}(\Delta\Omega)
=\cos^{2}\!\frac{\theta}{2}\,{\cal F}_{N}(\Delta \Omega )
+\sin^{2}\!\frac{\theta}{2}\,{\cal F}_{M}(\Delta \Omega ),
\end{equation}
with no cross term because tracing over the spacetime branches removes coherence.

Figure~\ref{PE_vs_theta} directly illustrates this $\theta$-dependence. Panels (a) and (b) show the coherent and incoherent probabilities, panel (c) shows their difference $\mathscr{P}_{\mathrm{cl}}-\mathscr{P}$, and panel (d) shows the ratio $\mathscr{P}/\mathscr{P}_{\mathrm{cl}}$. The separation is largest near $\theta=\pi/2$, where both branches have appreciable weight, and vanishes at $\theta=0,\pi$, where one branch amplitude is zero and hence $\mathscr{P}=\mathscr{P}_{\mathrm{cl}}$.

The phase in the angle scan is constructive: for the plotted parameters, ${\cal L}_{NM}>0$ and $\cos(\Delta\omega\,\mathcal{T})>0$. Nevertheless, constructive interference does not require $\mathscr{P}>\mathscr{P}_{\mathrm{cl}}$, because the coherent expression contains squared branch probabilities whereas the mixed state result is a convex sum. Writing $a=\cos^2(\theta/2)$, $b=\sin^2(\theta/2)$, and $\Phi=\Delta\omega\mathcal{T}$ gives
\begin{equation}
\mathscr{P}_{\mathrm{cl}}-\mathscr{P}
=ab\left({\cal F}_N+{\cal F}_M-2\cos\Phi\,{\cal L}_{NM}\right)\geq0,
\end{equation}
where the inequality follows from $|{\cal L}_{NM}|\leq\sqrt{{\cal F}_N{\cal F}_M}\leq({\cal F}_N+{\cal F}_M)/2$. Thus the constructive phase minimizes the deficit; a destructive relative phase would increase it. The deficit also vanishes when the geometries coincide ($N=M$), since then ${\cal L}_{NN}={\cal F}_N$, and at the single-branch endpoints $\theta=0,\pi$.

\begin{figure}[ht]
\centering
     \includegraphics[width=\linewidth]{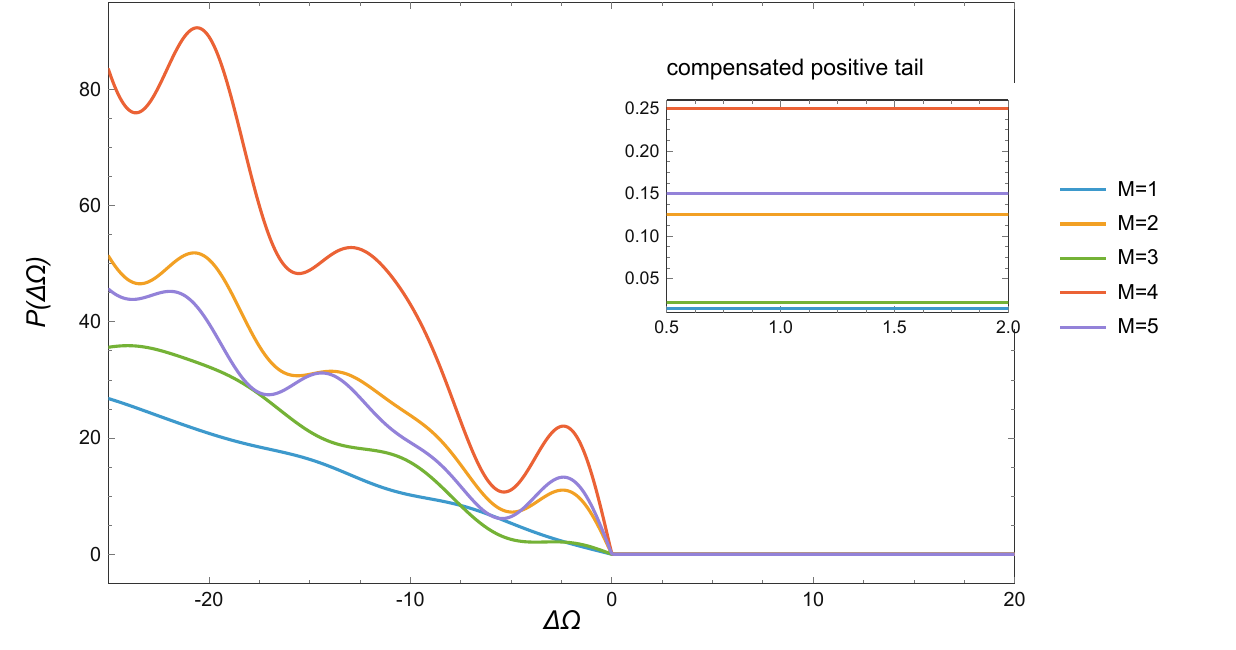}
          \caption{Excitation transition probability $\mathscr{P}(\Delta\Omega)$ for positive and negative detector gap, evaluated for $\mathcal{T}=30$, $r=0.5$, $\theta=\pi/2$, and $N=4$. Here the horizontal variable denotes signed transition energy: positive values describe excitation from the ground state, while negative values describe de-excitation from an initially excited detector. The inset plots the compensated positive-gap tail $e^{\Delta\Omega\mathcal{T}}\mathscr{P}$ over $0.5\leq\Delta\Omega\leq2$; its horizontal lines expose the asymptotic coefficients and the order $M_2<M_1<M_3<M_4<M_5$. The coincident case $M_4=N$ recovers the single-string return response. The five cases are distinguished by the same solid-curve colour coding used in the preceding figures.}
          \label{PE_vs_omega_T30_r0.1_theta_pihalf_N4}
\end{figure}

\section{Resonant Topological Matching}

A particularly sharp signature of geometric interference appears when scanning the ratio of topological charges. As shown in Fig.~\ref{PE_vs_gamma_T0.01_omega1_r1_theta_pihalf_N1to100_M10}, the excitation probability exhibits a dominant resonance at $M=N$ and a sequence of secondary maxima at rational ratios of the topological charges. This pattern is a direct consequence of the discrete image structure in the cross-branch correlator, which depends on $\sin^2[\pi(n/N-m/M)]$ and therefore becomes enhanced when the two image lattices align.

Figure~11 shows the excitation probability $\mathscr{P}(\Delta\Omega)$ as a function of the detector energy gap for fixed $N=4$ and varying $M$, evaluated at $r=0.5$, $\theta=\pi/2$, and $\mathcal{T}=30$. Excitations ($\Delta\Omega>0$) are strongly suppressed for long interaction times, while de-excitations ($\Delta\Omega<0$) exhibit an oscillatory decay whose amplitude depends on the topological mismatch between the branches. Dividing the positive-gap tail by its common factor $e^{-\Delta\Omega\mathcal{T}}$ yields the horizontal inset plateaus, which show that the asymptotic ordering is $M_2<M_1<M_3<M_4<M_5$ with no crossing. The case $M_4=N$ recovers the single-string return response.

\section{Conclusions and further discussion}

We have investigated an operational signature of quantum spacetime superposition using an inertial Unruh-DeWitt detector coupled to a massless scalar field on a coherent superposition of two conical (cosmic string) geometries. Each branch is characterized by an integer topological charge, $N$ or $M$, and the spacetime state
\begin{equation}
|\mu\rangle=\cos(\theta/2)\,|N\rangle+e^{i\phi}\sin(\theta/2)\,|M\rangle
\end{equation}
is naturally parametrized by the Bloch-sphere angle $\theta$, which controls the relative weight of the two geometric branches. In this language, $\theta=0,\pi$ correspond to single-geometry configurations, while $\theta=\pi/2$ represents a balanced superposition and maximizes the potential for geometric interference.

Using the method of images, we constructed the Wightman functions for individual cosmic string spacetimes and introduced a symmetric, well-defined ansatz for the cross-branch correlator that reduces to the standard result when $N=M$ and to Minkowski spacetime when $N=M=1$. With a Lorentzian switching of width $\mathcal{T}$, this yields closed-form expressions for the branch response functions ${\cal F}_{N,M}(E)$ and the interference response ${\cal L}_{NM}(E)$, allowing a systematic comparison between coherent (quantum) superpositions and incoherent (classical) mixtures of geometries.

A central result is that the conditional excitation probability of the detector contains a genuinely quantum interference term proportional to
\begin{equation}
\sin^{2}\theta \;\cos(\Delta\omega\,\mathcal{T})\,{\cal L}_{NM}(\Delta\Omega),
\end{equation}
together with a geometric post-selection normalization factor. No such term appears in the response for the  (unconditional) mixed case, which is simply the convex sum
\begin{equation}
\mathscr{P}_{\mathrm{cl}}=\cos^{2}(\theta/2)\,{\cal F}_{N}+\sin^{2}(\theta/2)\,{\cal F}_{M}.
\end{equation}
The difference between quantum and mixed state predictions therefore vanishes when either branch amplitude is zero, when the branches are identical ($N=M$), or in regimes where detector excitations are suppressed altogether (large gap or long interaction time).

Across the parameter ranges explored in our figures, the unnormalized
coherent response $\mathscr{P}$ is no larger than the  
mixture $\mathscr{P}_{\mathrm{cl}}$. The phase convention used in the
plots is constructive: for the plotted parameters,
$\cos(\Delta\omega\mathcal{T})>0$ and ${\cal L}_{NM}>0$, so the cross
term increases $\mathscr{P}$ relative to its value at a destructive
phase. Nevertheless, the squared branch weights in the coherent
response imply
\[
\mathscr{P}_{\mathrm{cl}}-\mathscr{P}
=ab\left({\cal F}_N+{\cal F}_M
-2\cos\Phi\,{\cal L}_{NM}\right)\geq0.
\]
Constructive interference therefore minimizes the suppression relative
to the classical mixture, whereas a destructive phase increases it.

Our parameter scans further show that (i) excitation probabilities decay rapidly with increasing detector gap, (ii) finite-time switching is essential for resolving interference effects, (iii) the detector-string distance $r$ controls the strength of image-induced corrections, and (iv) resonant enhancement occurs when the image structures of the two branches align ($N=M$ or simple rational ratios). Taken together, these results demonstrate that coherence between spacetime geometries leaves a measurable imprint on a simple detector model that cannot be replicated by (classical) stochastic mixtures of geometries.

Our analysis relies on controlled approximations: second-order perturbation theory in the detector-field coupling, inertial trajectories, a massless scalar field, and a common reference vacuum for the two conical branches. While a more complete treatment could incorporate inequivalent Fock spaces or geometric backreaction, none of these refinements are expected to alter the qualitative conclusion. Interference between spacetime branches—parametrized by $\theta$ and the relative branch phase—constitutes a robust, operational signature of quantum spacetime superposition.

\section*{Acknowledgements}
This work was supported in part by the Nature Sciences and Engineering Research Council of Canada.

\bibliography{biblio}

\end{document}